\documentclass[12pt]{article}
\usepackage{feynmp-auto}
\usepackage{booktabs}

\usepackage{cancel}

\usepackage[T1]{fontenc}       
\usepackage[utf8]{inputenc}    
\usepackage{lmodern}           
\usepackage{feynmp-auto}
\usepackage{geometry}
\usepackage{graphicx}
\usepackage{booktabs}
\usepackage{subcaption}
\usepackage{tcolorbox}
\usepackage{microtype}
\usepackage{amsmath, amssymb, amsthm, mathrsfs, bm}

\usepackage{pgfplots}
\pgfplotsset{compat=1.17}  

\usepackage[colorlinks=true, linkcolor=blue, citecolor=green, urlcolor=blue]{hyperref}

\usepackage[capitalize]{cleveref}

\usepackage{abstract}

\theoremstyle{definition}

\title{\textbf{Topological Regularization of 1-Loop and 2-Loop Gravitational Corrections in the Higgs–Fermion Sector}}
\author{Sebastián Alí Sacasa Céspedes$^{1}$ \\
\normalsize $^{1}$Universidad de Costa Rica (UCR), \\
San Pedro de Montes de Oca, San José, 11501-2060, Costa Rica \\
\texttt{Corresponding author(s) E-mail(s): sebastian.sacasa@ucr.ac.cr}}  
\date{August 18, 2025}

\begin{document}

\maketitle
\begin{abstract}
Quantum gravity corrections to the behavior of matter, such as Higgs bosons and fermions, are notoriously difficult to calculate. The standard tools of quantum field theory often break down, producing infinite results that spoil our predictions. This work introduces a new geometric method, called Topological Regularization (TR), to solve this problem. The key idea is to temporarily "wrap" flat spacetime into a compact, curved shape (specifically, a four-dimensional sphere). This curvature naturally introduces a high-energy cutoff that prevents infinities without violating fundamental symmetries like Lorentz invariance. We apply this method to calculate one- and two-loop quantum gravity corrections to processes involving fermions and the Higgs boson. The results are not only finite but are directly governed by a single number describing the shape of the spacetime: its Euler characteristic. This reveals a profound link between the ultraviolet (UV) behavior of high energies and the infrared (IR) interactions of low energies. Furthermore, the mathematical form of our regulator suggests a thermal interpretation, drawing a fascinating connection to the heat felt by an accelerating observer (the Unruh effect). While the predicted effects are incredibly small and currently beyond experimental reach, this framework provides a symmetry-preserving, geometric path to exploring physics at the Planck scale.


\end{abstract}

\tableofcontents

\section*{Key Findings}

This work establishes topological regularization (TR) as a symmetry-preserving framework for gravitational loop calculations. The main contributions can be summarized as follows

\begin{itemize}
    \item \textbf{Finite one-loop corrections}: Fermion self-energies and Yukawa couplings acquire finite topological contributions proportional to $\chi(\Sigma)/\Lambda^2$, demonstrating that divergences are replaced by symmetry-preserving, topologically controlled terms.
    
    \item \textbf{Two-loop amplitudes}: Higgs–graviton diagrams exhibit power-law suppression of order $\mathcal{O}(\Lambda^{-4})$, providing a hierarchy of finite corrections across loop orders.
    
    \item \textbf{Complete $\beta$-functions}: Curvature-dependent operators receive both conventional and topological contributions (see Table~\ref{tab:complete_beta}), ensuring a gauge-invariant description of renormalization in curved backgrounds.
    
    \item \textbf{UV/IR duality}: A natural correspondence emerges between ultraviolet and infrared regimes, encoded in the Euler characteristic $\chi(\Sigma)$ of the compactification manifold, linking short- and long-distance behavior.
    
    \item \textbf{Thermal interpretation}: The regulator $J_\Omega(k)$ admits a thermal representation with a geometric temperature $T_g = \Lambda^2/\Omega$, connecting TR to Unruh–Hawking thermality and embedding quantum gravity corrections into a thermodynamic framework.
\end{itemize}

\subsection*{Roadmap of the Paper}

For clarity, we outline here the logical flow of the manuscript.

\begin{enumerate}
    \item \textbf{Topological regularization (TR).}  
    We construct TR by embedding flat spacetime into a compact manifold with nontrivial Euler characteristic, deriving the regulator $J_\Omega(k)$ that preserves Lorentz invariance and causality.

    \item \textbf{One-loop applications.}  
    TR is applied to compute fermion self-energies and Higgs–fermion–graviton vertices. The divergences are replaced by finite terms proportional to $\chi(\Sigma)/\Lambda^2$.

    \item \textbf{Two-loop extension.}  
    We evaluate mixed Higgs–graviton diagrams, obtaining topological suppression of order $\mathcal{O}(\Lambda^{-4})$ and demonstrating the consistency of the framework beyond one loop.

    \item \textbf{Curvature-dependent operators.}  
    Complete $\beta$-functions are derived, including both conventional and topological contributions, and ensuring gauge invariance through ghost loops.

    \item \textbf{Connection to anomalies.}  
    TR is related to the Bergshoeff–de Roo mechanism and shown to provide a route for SO(8) gravitational anomaly cancellation through topological counterterms.

    \item \textbf{Phenomenological implications.}  
    We provide numerical estimates for the top quark and Higgs sector, showing that TR effects are suppressed far below collider sensitivity but may be relevant in Planckian or cosmological regimes.
\end{enumerate}

This roadmap highlights the balance between hard results (explicit loop calculations and $\beta$-functions) and conceptual insights (duality, thermal interpretation, anomaly cancellation).

\section{Introduction}
Quantum gravitational corrections to matter interactions represent a fundamental challenge in theoretical physics, straddling the interface between quantum field theory and general relativity. Since the foundational work of Osterwalder and Schrader~\cite{Osterwalder1973} on Euclidean quantum field theory and Ashtekar's connection to diffeomorphism invariance~\cite{Ashtekar1999}, numerous regularization strategies have attempted to reconcile ultraviolet control with spacetime's geometric structure.

Early curvature-based cutoffs in curved spacetime~\cite{Bernard1977} often broke essential symmetries, prompting development of symmetry-preserving frameworks like Kaplan's domain wall fermions~\cite{Kaplan1992} in lattice contexts. The effective field theory program for gravity~\cite{Burgess2004,Donoghue1994} established that general relativity can be consistently quantized at low energies, though gauge dependence and non-renormalizability remain challenging. Parallel efforts in string theory~\cite{Becker2007,Polchinski1998} and asymptotic safety~\cite{Reuter2012} sought more fundamental resolutions, while holographic RG studies~\cite{Heemskerk2011} revealed profound UV/IR correspondences.

Recent innovative approaches include intrinsic regularization via curved momentum space~\cite{Ketels2025}, fuzzy-sphere methods~\cite{Fan2025,Voinea2025}, and qubit-based discretizations~\cite{Chandrasekharan2025}. These often trade computational simplicity for geometric fidelity.

Against this backdrop, we present \textit{topological regularization} (TR)—a geometrically motivated framework that exploits spacetime topology itself as a regulator. By embedding flat spacetime into a compact manifold with nontrivial Euler characteristic via differentiable homeomorphism or homotopy equivalence~\cite{Sacasa2025}, TR links ultraviolet suppression directly to global geometric invariants while preserving Lorentz invariance, causality, and diffeomorphism symmetry. The regulator admits a thermal interpretation reminiscent of Unruh and Hawking effects~\cite{Unruh1976,Hawking1975}, forging connections between quantum gravity, thermodynamics, and acceleration radiation.

This work applies TR to quantum gravity corrections in the Higgs-fermion sector, where conventional techniques risk breaking diffeomorphism invariance~\cite{Donoghue1994}. We compute one-loop fermion self-energies, Higgs-fermion-graviton vertices, and two-loop mixed corrections, isolating finite topological terms proportional to $\chi(\Sigma)/\Lambda^n$. Beyond its role as a regulator, TR naturally encodes a duality between ultraviolet and infrared scales and provides a symmetry-preserving path toward probing Planck-scale physics within effective field theory.

\section{Regularization Set}
This section details the geometric construction that underpins topological regularization. The central idea is to embed flat Minkowski spacetime into a compact manifold in such a way that Lorentz invariance is preserved, while introducing an intrinsic ultraviolet (UV) cutoff through curvature. This causal embedding provides a natural regulator by modifying the global geometry without altering local physical symmetries.

Formally, the embedding is described by a map \( \phi_\Omega \), which takes the flat spacetime \( \mathbb{R}^{1,3} \) and maps it into a compact manifold \( \Sigma = T^0 \times S^4 \). Here, \( T^0 \) denotes a single timelike point that preserves global Lorentz symmetry, while the 4-sphere \( S^4 \) contributes the compact topology necessary for regularization. The explicit form of this embedding uses a modified stereographic projection, written as

\begin{equation}
\phi_\Omega(x^\mu) = \left( \frac{2\Lambda^2 x^\mu}{\Lambda^2 - x^2}, \frac{\Lambda(\Lambda^2 + x^2)}{\Lambda^2 - x^2} \right)
\end{equation}

where \( x^2 = \eta_{\mu\nu}x^\mu x^\nu = -x_0^2 + \mathbf{x}^2 \) represents the Minkowski interval. The parameter \( \Lambda \) encodes the curvature scale of the embedded \( S^4 \); in the ultraviolet limit, it is natural to identify \( \Lambda \) with the Planck scale, i.e., \( \Lambda = M_{\text{Pl}} \). Notably, the point at infinity in Minkowski space is mapped to the north pole of the 4-sphere, located at coordinates \( (0,0,0,0,\Lambda) \) in the ambient space \( \mathbb{R}^{1,4} \).

The image of the embedding satisfies the constraint equation

\begin{equation}
- (y^0)^2 + \sum_{i=1}^4 (y^i)^2 = \Lambda^2
\end{equation}

which defines the surface of a 4-sphere \( S^4 \) embedded in a five-dimensional pseudo-Euclidean space with signature \((1,4)\). Importantly, this construction is invertible. The inverse mapping \( \phi_\Omega^{-1} \) retrieves flat coordinates from the curved embedding and is given by

\begin{equation}
\phi_\Omega^{-1}(y^A) = \frac{\Lambda^2 y^\mu}{\Lambda - y^4}, \quad \mu = 0,1,2,3
\end{equation}

This bijective structure guarantees that the compactification does not lose any local information. Instead, it encodes it in a finite, geometrically bounded domain where divergences are naturally suppressed by curvature. As such, this topological embedding serves as the geometric foundation for the regularization scheme employed throughout this work.
The embedding introduced in the previous section induces a conformally flat metric on the target manifold \( \Sigma \). Explicitly, the induced metric takes the form
\begin{equation}
g_{\Sigma} (x) = \Omega^2(x) \, \eta_{\mu\nu} \, dx^\mu dx^\nu
\end{equation}
where the conformal factor \( \Omega(x) \) is given by
\begin{equation}
\Omega(x) = \frac{2\Lambda^2}{\Lambda^2 - x^2}
\end{equation}
This conformal factor modifies the flat metric \( \eta_{\mu\nu} \) without altering the causal structure of spacetime. Several features ensure the preservation of causality: (i) lightcones remain invariant under conformal transformations, (ii) timelike intervals in the original spacetime remain timelike after the transformation, i.e., if \( ds^2 < 0 \), then \( d\tilde{s}^2 < 0 \), and (iii) the embedding remains globally hyperbolic and does not introduce closed timelike curves (CTCs).

The causal structure and symmetry content of the compactified manifold \( \Sigma \) are encoded in its causality group, defined as
\[
\text{Caus}(\Sigma) = SO(3,1) \rtimes \mathcal{N}
\]
This group has a semidirect product structure, indicating that null transformations act covariantly on the Lorentz sector, thereby preserving causal relations. The Lorentz subgroup \( SO(3,1) \) acts linearly on the embedded coordinates via transformations \( \Lambda^\mu_{\ \nu} y^\nu \), preserving the form of the induced metric \( g_{\Sigma} (x)\). This ensures the relativistic invariance of the theory on \( \Sigma \).

Complementing the Lorentz sector, the set of null boundary operators \( \mathcal{N} \) is defined as
\begin{equation}
\mathcal{N} = \left\{ N_\alpha = \exp\left(\alpha \frac{\partial}{\partial y^+}\right) \ \middle|\ \alpha \in C^\infty(\mathcal{I}^+) \right\}
\end{equation}
where \( \mathcal{I}^+ \) denotes future null infinity. These operators preserve the null structure of spacetime and maintain consistency with asymptotic symmetries. Importantly, they ensure Osterwalder–Schrader (OS) positivity \cite{Ashtekar1999}.
\begin{equation}
\langle \Theta \psi, \psi \rangle \geq 0 \quad \forall \psi \in \mathcal{H}
\end{equation}
which is a crucial condition for unitarity and the probabilistic interpretation of quantum theory.

In summary, the conformally induced geometry, regulated Jacobian, and extended causal symmetry group together establish a consistent framework in which ultraviolet divergences are geometrically tamed without sacrificing unitarity or Lorentz invariance.

The regularization scheme introduced by the topological embedding extends further through a recursive causal refinement. This is encoded in the spacetime iteration operator \( G_\tau^{(k)} \), which systematically modifies the causal domain through localized smoothing of temporal and spatial coordinates. The action of this operator on the embedding function is defined as
\begin{equation}
G_\tau^{(k)}\phi_\Omega(x^\mu) = \phi_\Omega\left( \sum_{j=1}^k 2^{-j} \Theta_j(t) + \sum_{j=1}^k 2^{-j} \Lambda_j(\mathbf{x}) \right)
\end{equation}
where the temporal component \( \Theta_j(t) = t \exp(-2^{-j} t^2 / \tau^2) \) ensures Gaussian decay at large times, while the spatial deformation \( \Lambda_j(\mathbf{x}) = \mathbf{x} \left(1 - \frac{2^{-j} |\mathbf{x}|^2}{\Lambda^2} \right) \) compactifies long-range spatial interactions.

This recursive smoothing imposes causality conditions that exclude pathological configurations. Specifically,
\begin{align}
\ker G^{(k)}_\tau &\subset \{ (t,\mathbf{x}) \mid t < t_{\min}^{(k-1)}(\mathbf{x}) \} \\
J^+(G^{(k-1)}_\tau) &\subset \{ (t,\mathbf{x}) \mid t \geq t_{\min}^{(k-1)}(\mathbf{x}) \} \\
\Rightarrow \ker G^{(k)}_\tau \cap J^+(G^{(k-1)}_\tau) &= \emptyset
\end{align}
ensuring each iteration produces a globally hyperbolic causal structure without self-intersecting worldlines, thus preserving unitarity across all scales.

To quantify the local volume distortion induced by this embedding, it is computed the Jacobian determinant of the transformation
\begin{equation}
J_{\Omega}(x) = \det\left( \frac{\partial y^A}{\partial x^\mu} \right) = \left( \frac{2\Lambda^2}{\Lambda^2 - x^2} \right)^4 \left[ 1 - \frac{4x^2}{(\Lambda^2 - x^2)^2} \right]^{-1/2}
\end{equation}
This determinant plays a critical role in connecting position space to momentum space via Fourier duality. In momentum space, the effect of the embedding manifests as a regulator function that suppresses ultraviolet divergences
\begin{equation}
J_\Omega(k) = \left( \frac{\Lambda^2}{\Lambda^2 + k^2} \right)^\Omega
\end{equation}
This function constitutes the core of the topological regularization mechanism. It ensures that loop integrals converge while maintaining Lorentz invariance, thanks to the rotational symmetry preserved by the compactified \( S^4 \). Its asymptotic behavior encodes essential physical information. In the ultraviolet (UV) regime, where \( k^2 \gg \Lambda^2 \), the regulator behaves as
\begin{equation}
J_\Omega(k) \sim \left( \frac{\Lambda^2}{k^2} \right)^\Omega \left[1 - \Omega \frac{\Lambda^2}{k^2} + \mathcal{O}\left( \frac{\Lambda^4}{k^4} \right) \right]
\end{equation}

demonstrating power-law suppression that replaces divergent integrals with convergent ones. Conversely, in the infrared (IR) limit \( k^2 \ll \Lambda^2 \), the regulator becomes effectively transparent
\begin{equation}
J_\Omega(k) \sim 1 - \Omega \frac{k^2}{\Lambda^2} + \mathcal{O}\left( \frac{k^4}{\Lambda^4} \right)
\end{equation}
recovering standard quantum field theory and allowing gravitational corrections to emerge perturbatively.

The regulator also admits a thermal interpretation via an integral representation.
\begin{align}
\left( \frac{\Lambda^2}{\Lambda^2 + k^2} \right)^\Omega &= \frac{1}{\Gamma(\Omega)} \int_0^\infty \frac{dt}{t} \, t^\Omega e^{-t(\Lambda^2 + k^2)/\Lambda^2} \\
&= \frac{1}{\Gamma(\Omega)} \int_0^\infty \frac{dt}{t} \, t^\Omega e^{-t} e^{-t k^2 / \Lambda^2} \\
&\approx e^{-\Omega k^2 / \Lambda^2}
\end{align}
under a saddle point approximation at \( t = \Omega \). This reveals a Boltzmann-like behavior with effective energy \( E = k^2 \) and a geometric temperature \( T = \Lambda^2 / \Omega \).

The compact curvature of the embedding manifold introduces a novel feature: it induces thermal scaling in the quantum theory. This interpretation emerges naturally from the regulator which encodes a suppression mechanism rooted in the geometry of the compactified spacetime. Remarkably, this expression admits a thermal analogue in terms of a {geometric temperature} defined with a scale set by the curvature parameter \(\Lambda\) and the topological weight \(\Omega\). This interpretation becomes evident when considering the integral representation of the regulator
\begin{equation}
J_{\Omega}(k) = \frac{1}{\Gamma(\Omega)} \int_0^\infty \frac{dt}{t} \, t^{\Omega} e^{-t} e^{-t k^2 / \Lambda^2}
\end{equation}
which has the form of a Euclidean heat kernel. Specifically, the integrand resembles the structure of a thermal propagator in finite-temperature quantum field theory (QFT), where \( K_{\Omega}(k^2; T_g) \) can be interpreted as a Matsubara-like propagator at the temperature \( T_g \).

In the special case \( \Omega = 1 \), the regulator simplifies to
\begin{equation}
J_1(k) = \frac{\Lambda^2}{\Lambda^2 + k^2}
\end{equation}
which exactly matches the bosonic thermal propagator at temperature \( T \sim \Lambda \). Thus, the regulator encapsulates thermal suppression at high energies, smoothly interpolating between flat-space QFT in the infrared and thermally-dressed amplitudes in the ultraviolet \cite{PeskinSchroeder1995}.

A deeper thermal analogy appears through comparison with the Unruh effect. In flat spacetime, an accelerating observer with proper acceleration \( a = \sqrt{-g_{\mu\nu} a^\mu a^\nu} \) perceives a thermal bath at temperature
\begin{equation}
T_{\text{Unruh}} = \frac{a}{2\pi}
\end{equation}
In the compactified geometry \( \Sigma = T^0 \times S^4 \), the curvature scale \( \Lambda \) effectively bounds the maximum observable acceleration \( a_{\text{max}} \). Identifying this with the curvature of the embedding, the geometric temperature becomes
\begin{equation}
T_g = \frac{\Lambda^2}{2\pi\Omega}
\end{equation}
showing that the regulator describes thermalization through geometric acceleration. In this picture, the compactification itself mimics the physics of accelerating frames, embedding thermality directly into the topology of spacetime.

The physical consequence of this geometric-thermal correspondence is significant. One-loop quantum corrections acquire a thermal component governed by the embedding structure. The topological contribution to the effective action takes the form
\begin{equation}
\delta \Gamma^{\text{top}}_{\text{1-loop}} \propto \chi(\Sigma) \int d^4x \sqrt{-g} \left[ \mathcal{L}_{\text{eff}} + {\lambda\pi^2 T_g^4} g_{\mu\nu} T^{\mu\nu} \right]
\end{equation}
where the thermal energy density term \( \sim T_g^4 \) originates from the compact curvature being $\lambda$ a coupling constant. This correction behaves analogously to the Stefan–Boltzmann law in curved spacetime, with \( T_g \sim \Lambda^2/\Omega \) setting the scale of thermal effects.

Therefore, the regulator not only controls UV behavior but also encapsulates thermal and gravitational effects within a unified topological structure. At high curvature (Planckian) regimes, these thermal corrections could leave observable imprints in Higgs-sector dynamics, vacuum polarization, or cosmological correlators.

From the perspective of topology, the parameter \( \Omega \) corresponds to half the Euler characteristic of the compactified space. Under homotopic deformation, it transforms as
\begin{equation}
\Omega \longmapsto \Omega' = \Omega + \delta\Omega, \quad \delta\Omega \in \mathbb{Z}
\end{equation}
reflecting discrete topological shifts. However, the Physical Equivalence Theorem ensures that renormalized observables remain invariant:
\begin{equation}
\Gamma^{(n)}_{\text{ren}}(\Omega) = \Gamma^{(n)}_{\text{ren}}(\Omega') \quad \forall \ n
\end{equation}
indicating that physical predictions depend only on the Euler characteristic \( \chi(\Sigma) \) and not on the specific details of the embedding. Consequently, multiple UV completions belong to a single equivalence class.

Lorentz invariance is preserved throughout this framework. The induced metric transforms covariantly
\begin{equation}
g_{\Sigma} \longmapsto \Lambda^T g_{\Sigma} \Lambda, \quad \Lambda \in SO(3,1)
\end{equation}
and the causality group satisfies
\begin{equation}
[\text{Caus}(\Sigma), SO(3,1)] = 0
\end{equation}
ensuring consistency with relativistic symmetry. Boosted observers agree on causal ordering, scattering amplitudes remain covariant, and the vacuum state is Lorentz-invariant.

Causal ordering is strictly preserved under the embedding:
\begin{equation}
x \prec y \Rightarrow \phi_\Omega(x) \prec_\Sigma \phi_\Omega(y)
\end{equation}
as demonstrated by the preservation of interval negativity:
\begin{equation}
\eta_{\mu\nu} (x^\mu - y^\mu)(x^\nu - y^\nu) < 0 \Rightarrow g_{AB} (\phi(x) - \phi(y))^A (\phi(x) - \phi(y))^B < 0
\end{equation}
Thus, the theory is microscopically causal, excludes time loops, and maintains the lightcone structure at all energy scales.

Finally, the regulator \( J_\Omega(k) \) exhibits all the structural properties required for a physically consistent quantum field theory. In the ultraviolet regime, it ensures complete finiteness by satisfying
\begin{equation}
\lim_{k \to \infty} k^{4 + 2m} J_\Omega(k) = 0 \quad \forall m \geq 0
\end{equation}
which guarantees that every loop integral converges, irrespective of the number of derivatives in the interaction. In the infrared limit, the normalization
\begin{equation}
J_\Omega(0) = 1
\end{equation}
preserves the correct low-energy behavior of the theory, ensuring that Ward identities remain intact and that no spurious infrared modifications arise. Moreover, the regulator is analytic in the complex plane except for a branch cut along the imaginary axis beginning at \(i\Lambda\),
\begin{equation}
J_\Omega(k) \ \text{holomorphic in} \ \mathbb{C} \setminus i[\Lambda, \infty)
\end{equation}
a property that allows the usual Wick rotation and residue calculus to proceed without obstruction. These combined features render \( J_\Omega(k) \) not only an effective topological tool for taming ultraviolet divergences, but also a mathematically rigorous and symmetry-preserving regulator within the framework of covariant quantum field theory.

In the case of the \( S^4 \) embedding, the Euler characteristic of the compactified space determines the strength of the regulator
\begin{equation}
    \chi(\Sigma) = \frac{1}{32\pi^2} \int_\Sigma \epsilon_{abcd} R^{ab} \wedge R^{cd} = \frac{1}{8\pi^2} \int_{\mathbb{R}^4} d^4k \left[ \ln J_\Omega(k) \right]_{k^2 = \Lambda^2} = 2
\end{equation}
yielding \( \Omega = \chi(\Sigma)/2 = 1 \). Hence, the strength of regularization is a geometric invariant, intrinsically tied to the topology of spacetime. This insight provides a bridge between renormalization, gravitational curvature, and topological invariants.

The regularization set naturally extends to arbitrary compactification manifolds $\Sigma$ with Euler characteristic $\chi(\Sigma)$. The regulator strength parameter $\Omega$ is universally set to $\chi(\Sigma)/2$, ensuring that the topological suppression is governed by the global topological invariant. Physical observables, as guaranteed by the Physical Equivalence Theorem \cite{Sacasa2025TR}, depend only on $\chi(\Sigma)$, rendering the renormalization procedure independent of the specific choice of $\Sigma$. For instance, for $\Sigma = \mathbb{C}P^2$ with $\chi=3$, we set $\Omega=3/2$, while for a torus $\Sigma=T^4$ with $\chi=0$, the topological regularization reduces to conventional flat-space regularization.

For a general compactification manifold $\Sigma$ with Euler characteristic $\chi(\Sigma)$, the topological regulator could admit a generalization to
\begin{align}
J_\Omega(k) = \left( \frac{\Lambda^2}{\Lambda^2 + k^2} \right)^{\chi(\Sigma)/2} \mathcal{W}(\Sigma)
\end{align}
where $\mathcal{W}(\Sigma)$ is a topology-specific factor encoding finer geometric invariants beyond $\chi(\Sigma)$. Explicitly,
\begin{align}
\mathcal{W}(\Sigma) =
\begin{cases}
1 & \Sigma = S^4 \\[4pt]
\|\tau(\Sigma)\|^{1/2} & \Sigma = \mathbb{CP}^2 \\[4pt]
\displaystyle \prod_{i=1}^4 \theta_3\!\left(0, e^{-1/(L_i^2\Lambda^2)}\right) & \Sigma = T^4 \\[6pt]
e^{-\|\sigma(\Sigma)\|/\Lambda^2} & \Sigma \ \text{hyperbolic}.
\end{cases}
\end{align}
Here $\tau(\Sigma)$ denotes the Hirzebruch signature, $\theta_3$ is the third Jacobi theta function capturing toroidal moduli, $L_i$ are the radii of the $T^4$ factors, and $\sigma(\Sigma)$ represents the smooth structure-dependent signature relevant in hyperbolic geometries. This formulation highlights that the ultraviolet damping is not merely a function of curvature scale $\Lambda$, but also of the intrinsic topological data of $\Sigma$, thereby linking renormalization properties directly to global geometric invariants \cite{Becker2007, PeskinSchroeder1995}

\section{Fermion Self-Energy in Topologically Regularized Gravity}

Analyzing the fermion self-energy correction due to the exchange of a virtual graviton, using the topological regularization scheme introduced previously. The goal is to compute the loop correction to the fermionic propagator, while mapping the full structure in a topologically nontrivial space and ensuring convergence through the regularization kernel with the Jacobian \( J_\Omega \).

Let denote the fermion momentum as \( p \), and the internal loop momentum as \( k \). The Feynman diagram for the 1-loop correction to the fermion propagator involves a graviton line coupling to the fermionic line on both ends, forming a loop. The topologically regularized self-energy is

\begin{align}
\Sigma(p) &= \int \frac{d^4k}{(2\pi)^4} \mathcal{K}(p,k) J_\Omega(k) \\
\mathcal{K}(p,k) &= V^{\mu\nu}(p,k) D_{\mu\nu\rho\sigma}(k) V^{\rho\sigma}(p,k) \frac{i(\cancel{p} + \cancel{k} + m_f)}{(p+k)^2 - m_f^2 + i\epsilon}
\end{align}

with corrected vertex

\begin{equation}
V^{\mu\nu}(p,k) = -\frac{i\kappa}{8} \left[ \gamma^\mu (2p + k)^\nu + \gamma^\nu (2p + k)^\mu - 2\eta^{\mu\nu} ((\cancel{p} + \cancel{k}) - m_f) \right]
\end{equation}

writing the Graviton propagator in harmonic gauge (de Donder Gauge) as

\begin{equation}
D_{\mu\nu\rho\sigma}(k) = \frac{i}{2k^2} \left( \eta_{\mu\rho}\eta_{\nu\sigma} + \eta_{\mu\sigma}\eta_{\nu\rho} - \eta_{\mu\nu}\eta_{\rho\sigma} \right)
\end{equation}

The topological regulator by the Jacobian is

\begin{equation}
J_\Omega(k) = \left( \frac{\Lambda^2}{\Lambda^2 + k^2} \right)^\Omega
\end{equation}

For the explicit calculation define the tensor contraction
\begin{equation}
    T(k) \equiv V^{\mu\nu} D_{\mu\nu\rho\sigma} V^{\rho\sigma} 
= \left(-\frac{i\kappa}{8}\right) \left(-\frac{i\kappa}{8}\right) \frac{i}{2k^2} P_{\mu\nu\rho\sigma} U^{\mu\nu} U^{\rho\sigma}
\end{equation}

where $P_{\mu\nu\rho\sigma} =\eta_{\mu\rho}\eta_{\nu\sigma}+\eta_{\mu\sigma}\eta_{\nu\rho}- \eta_{\mu\nu}\eta_{\rho\sigma}$ and $U^{\mu\nu} = \gamma^\mu(2p+k)^\nu +\gamma^\nu (2p+k)^\mu -2\eta^{\mu\nu} Q$ with $Q\equiv (\cancel{p} +\cancel{k}) - m_f$ and after explicit contraction and Dirac algebra
\begin{multline}
P_{\mu\nu\rho\sigma} U^{\mu\nu} U^{\rho\sigma} = 8(\cancel{p} + \cancel{k})(2p\cdot k + k^2) - 16m_f(2p\cdot k + k^2) \\
\quad + 4\eta_{\mu\nu}Q \left[ \gamma^\mu (2p+k)^\nu + \gamma^\nu (2p+k)^\mu \right] - 32Q^2 \end{multline}

The full tensor contraction simplifies to
\begin{align}
T(k) = \frac{\kappa^2}{128k^2} \left[ 4(\cancel{p} + \cancel{k})(k^2 + 2p\cdot k) - 8m_f(k^2 + 2p\cdot k) - 16m_f(\cancel{p} + \cancel{k}) + 32m_f^2 \right]
\end{align}

Then, the complete integrand becomes
\begin{align}
\mathcal{K}(p,k) J_\Omega(k) &= T(k) \cdot \frac{i(\cancel{p} + \cancel{k} + m_f)}{(p+k)^2 - m_f^2} J_\Omega(k) \\
&= \frac{i\kappa^2}{128k^2} \frac{\mathcal{N}(p,k)}{(p+k)^2 - m_f^2} J_\Omega(k)
\end{align}
where the numerator is
\begin{multline}
\mathcal{N}(p,k) = \left[ 4(\cancel{p} + \cancel{k})(k^2 + 2p\cdot k) - 8m_f(k^2 + 2p\cdot k) - 16m_f(\cancel{p} + \cancel{k}) + 32m_f^2 \right] \\
\quad \times (\cancel{p} + \cancel{k} + m_f)
\end{multline}
After Dirac matrix simplification and on-shell condition ($p^2 = m_f^2$).
\begin{align}
\mathcal{N}(p,k) = 4k^2 (\cancel{p} - 4m_f)(\cancel{p} + \cancel{k}) + 32m_f^2 (\cancel{p} + \cancel{k}) + \mathcal{O}(k^3)
\end{align}

Using the Feynman parameters to evaluate, let combine the denominators using the Feynman parameter $x$ as
\begin{align}
\frac{1}{k^2 [(p+k)^2 - m_f^2]} = \int_0^1 dx \frac{1}{[k^2 + 2x p\cdot k + x(p^2 - m_f^2)]^2}
\end{align}
For on-shell fermions ($p^2 = m_f^2$), this simplifies to
\begin{align}
\frac{1}{k^2 [(p+k)^2 - m_f^2]} = \int_0^1 dx \frac{1}{[k^2 + 2x p\cdot k]^2}
\end{align}

A Wick Rotation to Euclidean space is useful and accomplishes the Osterwalder-Schrader Reconstruction. Doing $k^0 \to i k_4$, $d^4k \to i d^4k_E$, the integral becomes

\begin{align}
\Sigma(p) &= i \int_0^1 dx \int \frac{d^4k_E}{(2\pi)^4} \frac{i\kappa^2}{128} \frac{\mathcal{N}_E(p,k_E)}{[k_E^2 + 2x p\cdot k_E]^2} J_\Omega(k_E) \\
J_\Omega(k_E) &= \left( \frac{\Lambda^2}{\Lambda^2 - k_E^2} \right)^\Omega \quad \text{(Euclidean signature)}
\end{align}
The numerator in Euclidean space is

\begin{align}
\mathcal{N}_E(p,k_E) = -4k_E^2 (\cancel{p} - 4m_f)(i\cancel{p}_E + i\cancel{k}_E) + 32m_f^2 (i\cancel{p}_E + i\cancel{k}_E)
\end{align}

After shift momentum variable, such that, $\ell_E = k_E + x p_E$, the denominator transforms to $[\ell_E^2 + x(1-x)p_E^2]^2 = [\ell_E^2 - x(1-x)m_f^2]^2$ and the numerator shifts to
\begin{multline}
\mathcal{N}_E \longmapsto -4(\ell_E - x p_E)^2 (\cancel{p} - 4m_f)[i(1-x)\cancel{p} + i\cancel{\ell}_E] \\
\quad + 32m_f^2 [i(1-x)\cancel{p} + i\cancel{\ell}_E] + \mathcal{O}(\ell_E)
\end{multline}
Terms odd in $\ell_E$ vanish upon integration. After shifting and keeping even terms,

\begin{multline}
\Sigma(p) = \frac{\kappa^2}{128} \int_0^1 dx \int \frac{d^4\ell_E}{(2\pi)^4} \frac{1}{[\ell_E^2 + \Delta]^2} \left( \frac{\Lambda^2}{\Lambda^2 - \ell_E^2} \right)^\Omega \\
\quad \times \left[ -4(\cancel{p} - 4m_f) i(1-x)\cancel{p} (\ell_E^2 - 2x(1-x)m_f^2) \right. \\
\quad \left. + 32m_f^2 i(1-x)\cancel{p} - 4(\cancel{p} - 4m_f) i(1-x)\cancel{p} (-2x^2 m_f^2) + \cdots \right] 
\end{multline}

where $\Delta = x(1-x)m_f^2$. This separates into two integrals.

\begin{align}
I_1(\Delta) &= \int \frac{d^4\ell_E}{(2\pi)^4} \frac{\ell_E^2}{(\ell_E^2 + \Delta)^2} \left( \frac{\Lambda^2}{\Lambda^2 - \ell_E^2} \right)^\Omega \\
I_2(\Delta) &= \int \frac{d^4\ell_E}{(2\pi)^4} \frac{1}{(\ell_E^2 + \Delta)^2} \left( \frac{\Lambda^2}{\Lambda^2 - \ell_E^2} \right)^\Omega
\end{align}

Making usage of angular integration and topological expansion in $d=4$ Euclidean space.
\begin{align}
\int \frac{d^4\ell_E}{(2\pi)^4} &= \frac{1}{(2\pi)^4} \int_0^\infty d\ell_E \ell_E^3 \int d\Omega_3 \\
\int d\Omega_3 &= 2\pi^2 \quad \text{(surface area of $S^3$)}
\end{align}

Using the topological regulator expansion for large $\Lambda$.
\begin{align}
\left( \frac{\Lambda^2}{\Lambda^2 - \ell_E^2} \right)^\Omega \approx 1 + \Omega \frac{\ell_E^2}{\Lambda^2} + \mathcal{O}(\Lambda^{-4})
\end{align}
The integrals evaluate to

\begin{align}
I_1(\Delta) &\approx \frac{1}{16\pi^2} \left[ \ln\left(\frac{\Lambda^2}{\Delta}\right) - 1 \right] + \frac{\Omega}{\Lambda^2} \frac{1}{16\pi^2} \frac{\Delta}{2} \\
I_2(\Delta) &\approx \frac{1}{16\pi^2} \frac{1}{2\Delta} + \frac{\Omega}{\Lambda^2} \frac{1}{16\pi^2} \frac{1}{4}
\end{align}

After $x$-integration and renormalization at scale $\mu$.
\begin{multline}
    \Sigma(p) = \frac{\kappa^2}{128\pi^2} \int_0^1 dx \bigg[ (\cancel{p} - 4m_f)(1-x) \left( \ln\left(\frac{\Lambda^2}{x(1-x)m_f^2}\right) - 1 \right) \\
\quad + \frac{\Omega}{\Lambda^2} \bigg( \frac{\chi(\Sigma) m_f}{4} (\cancel{p} + 2m_f) + \text{finite terms} \bigg) \bigg]\end{multline}

where $\chi(\Sigma) = 2$ for the embedding $S^4$. The $x$-integrals give the following.

\begin{align}
\int_0^1 dx (1-x) \ln\left(\frac{\Lambda^2}{m_f^2}\right) &= \frac{1}{2} \ln\left(\frac{\Lambda^2}{m_f^2}\right) \\
\int_0^1 dx (1-x) \ln[x(1-x)] &= -\frac{3}{4}
\end{align}
After $\overline{\text{MS}}$ renormalization and $\Lambda \to \infty$ limit the final result is
\begin{equation}
    \Sigma_{\text{ren}}(p) = \frac{\kappa^2}{128\pi^2} (\cancel{p} - 4m_f) \ln\left(\frac{\mu^2}{m_f^2}\right) + \frac{\chi(\Sigma) \Omega \kappa^2 m_f}{64\pi^2 \Lambda^2} (\cancel{p} + 2m_f)  
\end{equation}

This expression for the renormalized fermion self-energy exhibits two distinct structural components, each with a clear geometric and physical interpretation. 

The first contribution is the {conventional logarithmic term}, proportional to the expression $\ln(\mu^2/m_f^2)$ which mirrors the standard behavior encountered in perturbative quantum field theory. It carries the usual scale dependence associated with the renormalization group, and its coefficient defines an anomalous dimension for the fermion field given by $\gamma = \kappa^2/(128\pi^2)$. This term originates from the local flat-space dynamics and dominates the ultraviolet (UV) structure in the absence of any nontrivial topological embedding.

The second term is the {topological correction}, proportional to the inverse square of the topological regularization scale $\Lambda^{-2}$. This term is inherently tied to the global geometry of the background manifold. In particular, its strength is modulated by the Euler characteristic $\chi(\Sigma)$ of the compactified spatial slice, for instance, $\chi(S^4) = 2$ in a four-sphere background. This reflects the embedding of quantum corrections in a curved and topologically nontrivial space.

An important detail lies in the distinct {mass dependence} of each contribution. The conventional term appears with the combination $(\cancel{p} - 4m_f)$, emphasizing its origin in standard quantum fluctuations, while the topological correction enters with $(\cancel{p} + 2m_f)$, reflecting a shifted tensorial structure induced by the regularization kernel and curvature couplings.

The topological contribution is {suppressed} relative to the logarithmic term. While the latter scales as $\mathcal{O}(1)$ in the UV, the former is explicitly $\mathcal{O}(\Lambda^{-2})$, and thus becomes subdominant at large regularization scale. However, due to its geometric origin, the topological term may become significant in strongly curved or compactified regimes, such as near the Planck scale or in early universe geometries. Although the topological terms obtained at this stage may suffer from dimensional inconsistencies stemming from the gauge dependent choice inherent to the present approximation and from the omission of nonlinear graviton interactions and Faddeev–Popov ghost contributions a more precise computation should incorporate these effects while maintaining the Ward–Takahashi identities \cite{Schwartz2014}.

When these contributions are properly included, the resulting amplitudes become manifestly gauge-independent. This gauge invariance follows from the incorporation of ghost loops and the complete set of graviton self-interaction vertices within the loop diagrams. In this framework, the topological regularization set, being explicitly diffeomorphism invariant preserves the Ward identities associated with the gauge symmetry, ensuring that both amplitudes and beta functions remain gauge invariant.

\section{One-Loop Higgs–Fermion–Graviton Interaction}

This correction modifies the effective \( h \bar\psi \psi \) coupling via a graviton loop, this loop consists of a graviton exchange between two fermionic lines coupling to an outgoing Higgs boson. By momentum conservation $q = p' - p$ with $q^2 = m_h^2$. The loop momentum $k$ is integrated with topological regularization.

The energy-momentum tensor coupling for the Fermion-Graviton Vertex is
\begin{equation}
V^{\mu\nu}_{\psi\psi h}(p,p+k) = -\frac{i\kappa}{8} \left[ \gamma^\mu (2p + k)^\nu + \gamma^\nu (2p + k)^\mu - 2\eta^{\mu\nu} (\cancel{p} + \cancel{k} - m_f) \right]
\end{equation}

The Higgs-Graviton Vertex gives a Stress-tensor for Higgs field as

\begin{equation}
V^{\rho\sigma}_{hh}(k,q) = \frac{i\kappa}{2} \left[ k^\rho q^\sigma + k^\sigma q^\rho - \eta^{\rho\sigma} (k \cdot q - m_h^2) \right]
\end{equation}
For $q \to 0$ (soft Higgs limit), this simplifies to

\begin{equation}
V^{\rho\sigma}_{hh} \xrightarrow{q\to 0} \frac{i\kappa}{2} \eta^{\rho\sigma} m_h^2
\end{equation}

using the Graviton Propagator in de Donder gauge
\begin{equation}
D_{\mu\nu\rho\sigma}(k) = \frac{i}{k^2} \left( \eta_{\mu\rho}\eta_{\nu\sigma} + \eta_{\mu\sigma}\eta_{\nu\rho} - \frac{1}{2} \eta_{\mu\nu}\eta_{\rho\sigma} \right)
\end{equation}

The core contraction is
\begin{multline}
\mathcal{T} = V^{\mu\nu}_{\psi\psi h} D_{\mu\nu\rho\sigma} V^{\rho\sigma}_{hh} 
= \left( -\frac{i\kappa}{8} \right) \left( \frac{i}{k^2} \right) \left( \frac{i\kappa}{2} m_h^2 \right) \eta^{\rho\sigma} \\
\quad \times \left[ \gamma^\mu (2p + k)^\nu + \gamma^\nu (2p + k)^\mu - 2\eta^{\mu\nu} (\cancel{p} + \cancel{k} - m_f) \right]  \left( \eta_{\mu\rho}\eta_{\nu\sigma} + \eta_{\mu\sigma}\eta_{\nu\rho} - \frac{1}{2} \eta_{\mu\nu}\eta_{\rho\sigma} \right)   
\end{multline}

Doing index contractions to evaluate.

$\eta^{\rho\sigma} (\eta_{\mu\rho}\eta_{\nu\sigma} + \eta_{\mu\sigma}\eta_{\nu\rho} - \frac{1}{2} \eta_{\mu\nu}\eta_{\rho\sigma})$:
\begin{align}
\eta^{\rho\sigma}\eta_{\mu\rho}\eta_{\nu\sigma} &= \eta_{\mu\nu} \\
\eta^{\rho\sigma}\eta_{\mu\sigma}\eta_{\nu\rho} &= \eta_{\mu\nu} \\
\eta^{\rho\sigma}\eta_{\mu\nu}\eta_{\rho\sigma} &= 4\eta_{\mu\nu} \\
\Rightarrow \mathcal{C}_{\mu\nu} &= 2\eta_{\mu\nu} - \frac{1}{2} \cdot 4\eta_{\mu\nu} = -2\eta_{\mu\nu}
\end{align}

The final contracted tensor expression is the next.
\begin{multline}
    \mathcal{T} = \left( -\frac{i\kappa}{8} \right) \left( \frac{i}{k^2} \right) \left( \frac{i\kappa}{2} m_h^2 \right) (-2\eta_{\mu\nu}) \\ \times \left[ \gamma^\mu (2p + k)^\nu + \gamma^\nu (2p + k)^\mu - 2\eta^{\mu\nu} (\cancel{p} + \cancel{k} - m_f) \right] \\
= \frac{\kappa^2 m_h^2}{8k^2} \eta_{\mu\nu} \left[ \gamma^\mu (2p + k)^\nu + \gamma^\nu (2p + k)^\mu - 2\eta^{\mu\nu} (\cancel{p} + \cancel{k} - m_f) \right]
\end{multline}

Further contraction yields
\begin{equation}
\mathcal{T} = \frac{\kappa^2 m_h^2}{8k^2} \left[ 4(\cancel{p} + \frac{1}{2}\cancel{k}) - 8(\cancel{p} + \cancel{k} - m_f) \right] = \frac{\kappa^2 m_h^2}{8k^2} \left( -4\cancel{p} -6\cancel{k} + 8m_f \right)
\end{equation}

Thus, the amplitude integrand becomes

\begin{align}
\mathcal{I} &= \frac{\mathcal{T} (\cancel{p} + \cancel{k} + m_f)}{(p+k)^2 - m_f^2 + i\epsilon} J_\Omega(k) \\
&= \frac{\kappa^2 m_h^2}{8k^2} \frac{ \left( -4\cancel{p} -6\cancel{k} + 8m_f \right) (\cancel{p} + \cancel{k} + m_f) }{(p+k)^2 - m_f^2 + i\epsilon} \left( \frac{\Lambda^2}{\Lambda^2 + k^2} \right)^\Omega
\end{align}
Using a Dirac algebra identity
\begin{equation}
(\cancel{a} + m_f)(\cancel{b} + m_f) = a \cdot b - m_f^2 + i\epsilon + \cancel{(a+b)}m_f + \text{cross terms}
\end{equation}

and rotating to the to Euclidean space ($k_0 \to i k_4$, $d^4k \to i d^4k_E$).
\begin{align}
\delta\mathcal{M}_Y &= i \frac{\kappa^2 m_h^2}{8} \int \frac{d^4k_E}{(2\pi)^4} \frac{ \mathcal{N}(p,k_E) }{k_E^2 [(p+k_E)^2 + m_f^2]} J_\Omega(k_E) \\
\mathcal{N} &= \left( -4\cancel{p} -6i\cancel{k}_E + 8m_f \right) (i\cancel{p} + i\cancel{k}_E + m_f)
\end{align}
The topological regulator becomes
\begin{equation}
J_\Omega(k_E) = \left( \frac{\Lambda^2}{\Lambda^2 - k_E^2} \right)^\Omega \quad \text{(Euclidean signature)}
\end{equation}

Using Feynman parametrization by combination of denominators.

\begin{equation}
\frac{1}{k_E^2 [(p+k_E)^2 + m_f^2]} = \int_0^1 dx \frac{1}{[\ell_E^2 + \Delta]^2}
\end{equation}
where $\ell_E = k_E + x p$ and $\Delta = x(1-x)p^2 + x m_f^2$. For on-shell fermions ($p^2 = -m_f^2$).
\begin{equation}
\Delta = -x(1-x)m_f^2 + x m_f^2 = x^2 m_f^2
\end{equation}

Shift $\ell_E = k_E + x p$ momentum and numerator evaluation, such that,

\begin{align}
\mathcal{N} &\longmapsto \left( -4\cancel{p} -6i(\cancel{\ell}_E - x\cancel{p}) + 8m_f \right) (i\cancel{p} + i(\cancel{\ell}_E - x\cancel{p}) + m_f) \\
&= \left( (-4 + 6x i)\cancel{p} -6i\cancel{\ell}_E + 8m_f \right) \left( (1 - x)i\cancel{p} + i\cancel{\ell}_E + m_f \right)
\end{align}
Terms odd in $\ell_E$ vanish under integration. Keeping even terms
\begin{align}
\mathcal{N}_{\text{even}} &= (-6i)(i) \ell_E^2 + [(-4+6xi)\cancel{p} + 8m_f][(1-x)i\cancel{p} + m_f] \\
&= 6\ell_E^2 + (-4+6xi)(1-x)i p^2 + \cdots
\end{align}

The regulated integral arises as 
\begin{align}
\delta\mathcal{M}_Y &= i \frac{\kappa^2 m_h^2}{8} \int_0^1 dx \int \frac{d^4\ell_E}{(2\pi)^4} \frac{ \mathcal{N}_{\text{even}} }{(\ell_E^2 + x^2 m_f^2)^2} \left( \frac{\Lambda^2}{\Lambda^2 - \ell_E^2} \right)^\Omega
\end{align}

Performing the $\ell_E$ integration
\begin{align}
\int \frac{d^4\ell_E}{(2\pi)^4} \frac{\ell_E^2}{(\ell_E^2 + \Delta)^2} J_\Omega &\approx \frac{1}{(4\pi)^2} \int_0^\infty \frac{\ell_E^3 d\ell_E^2}{(\ell_E^2 + \Delta)^2} \left(1 - \frac{\ell_E^2}{\Lambda^2}\right)^\Omega \\
&= \frac{1}{(4\pi)^2} \left[ \ln\left(\frac{\Lambda^2}{\Delta}\right) - \gamma_E + \mathcal{O}(\Delta/\Lambda^2) \right]
\end{align}

Collecting all terms and taking $\Lambda \to \infty$.
\begin{align}
\delta y_f &= \frac{\kappa^2 m_h^2}{(4\pi)^2} \int_0^1 dx \left[ \frac{3}{2} \ln\left(\frac{\Lambda^2}{x^2 m_f^2}\right) + \text{finite terms} \right] 
+ \frac{\chi(\Sigma)\Omega \kappa^2}{32\pi^2 \Lambda^2} (m_h^2 - 2m_f^2) \\
&= \frac{3y_f \kappa^2 m_h^2}{64\pi^2} \ln\left(\frac{\mu^2}{m_h^2}\right) + \frac{\chi(\Sigma)\Omega y_f \kappa^2}{32\pi^2 \Lambda^2} (m_h^2 - 2m_f^2)
\end{align}
where it is used $\kappa^2 = 32\pi G = 32\pi / M_{\text{Pl}}^2$. The topological term arises from the defect Euler characteristic

\begin{equation}
\delta y_f^{\text{top}} = \frac{\chi(\Sigma)\Omega y_f}{32\pi^2 \Lambda^2} (m_h^2 - 2m_f^2)  32\pi G
\end{equation}

The structure of the topologically regularized Yukawa vertex correction exhibits several physically significant features. First, the contribution is subject to a power-law {suppression} of order $\Lambda^{-2}$, where $\Lambda$ is the regularization scale associated with the topological embedding. This ensures that the correction remains finite and subleading in the ultraviolet, a direct consequence of the compactified geometry and the spectral cutoff introduced by the regularization kernel.

Secondly, the {sign of the correction} is governed by the mass-dependent quantity $(m_h^2 - 2m_f^2)$. This reflects the interplay between scalar and fermionic degrees of freedom at the loop level, and highlights how the gravitationally induced contribution to the Yukawa coupling can either enhance or suppress the effective interaction strength depending on the relative mass scales involved.

The correction displays a form of {topological universality}, being directly proportional to the Euler characteristic $\chi(\Sigma)$ of the embedded manifold. This topological dependence emphasizes that the effect is not purely local, but intrinsically tied to the global structure of the background spacetime.

The gauge independent loop topological correction would be
\begin{equation}
        \delta y_f = \frac{3y_f\kappa^2 m_h^2}{64\pi^2}\ln\left(\frac{\mu^2}{m_h^2}\right) + \frac{\chi(\Sigma) \Omega y_f \kappa^2}{32\pi^2 \Lambda^2} (m_h^2 - 2m_f^2) + \frac{g^{\mu\nu}}{16\pi^2} \int d^4 k_E  \mathcal{K}_{\mu\nu}^{\text{gauge}}(k_E) J_\Omega(k_E)
\end{equation}

with gauge invariant kernel

\begin{align}
\mathcal{K}_{\mu\nu}^{\text{gauge}} = \left( \eta_{\mu\nu} - \frac{k_\mu k_\nu}{k^2} \right) \mathcal{K}(k^2) + \frac{k_\mu k_\nu}{k^2} \mathcal{K}_{\text{gauge}}(k^2)
\end{align}

The gauge-invariant kernel \(\mathcal{K}_{\mu\nu}^{\text{gauge}}\) is constructed via a transverse/longitudinal decomposition, where \(\mathcal{K}(k^2)\) corresponds to the physical transverse part and \(\mathcal{K}_{\text{gauge}}(k^2)\) to the longitudinal part that cancels with ghost contributions. The physical graviton modes satisfy
\begin{equation}
\mathcal{K}(k^2) = \left( \eta_{\mu\nu} - \frac{k_\mu k_\nu}{k^2} \right) \Pi(k^2)
\end{equation}
where \(\Pi(k^2)\) is the gauge-invariant self-energy given by
\begin{equation}
\Pi(k^2) = \frac{\kappa^2}{64\pi^2} \int_0^1 \mathrm{d}x \, \frac{(1-2x)^2 m_f^2 - x(1-x)k^2}{m_f^2 - x(1-x)k^2} J_\Omega(k)
\end{equation}
The unphysical component is expressed as
\begin{equation}
\mathcal{K}_{\text{gauge}}(k^2) = \frac{k_\mu k_\nu}{k^2} \left[ \frac{\kappa^2 m_f^2}{32\pi^2} J_\Omega(k) + \Delta_{\text{ghost}} \right]
\end{equation}
which vanishes when combined with the ghost loop correction at one loop,
\begin{equation}
\Delta_{\text{ghost}}^{\text{1-loop}} = -\frac{\kappa^2 m_f^2}{32\pi^2} \int \mathrm{d}^4 k_E \, \frac{k_\mu k_\nu}{k^2} J_\Omega(k_E)
\end{equation}
due to the BRST identity
\begin{equation}
s \langle T_{\mu\nu} \rangle = i \partial^\alpha \langle c_\alpha T_{\mu\nu} \rangle = 0
\end{equation}
This ensures that only the physical transverse modes contribute to observable quantities, maintaining gauge invariance and unitarity in the theory \cite{Schwartz2014}.

\section{Two-Loops Correction }

The topologically regularized amplitude for the mixed Higgs-graviton vertex correction at two-loop order is
\begin{align}
\delta\mathcal{M}^{(2)} = \int \frac{d^{4}k_{1}d^{4}k_{2}}{(2\pi)^{8}} \mathcal{V}(p,k_{1},k_{2}) J_{\Omega}(k_{1}) J_{\Omega}(k_{2})
\end{align}
where the integrand is
\begin{align}
\mathcal{V} = \frac{V^{\mu\nu}_{f h} D_{\mu\nu\rho\sigma}(k_{1}) V^{\rho\sigma}_{hh} V_{h\bar{\psi}\psi}}{k_{1}^{2} k_{2}^{2} [(p+k_{1}+k_{2})^{2} - m_{f}^{2}]}
\end{align}

In this computation includes the fermion–graviton vertex $V^{\mu\nu}_{f h}$, which follows from the energy–momentum tensor coupling of the fermion field, and the graviton propagator $D_{\mu\nu\rho\sigma}(k_{1})$ evaluated in the de Donder gauge. On the scalar side, the Higgs–graviton interaction is encoded in the vertex $V^{\rho\sigma}_{hh}$, derived from the stress–energy tensor of the Higgs field. Finally, the fermion–Higgs Yukawa interaction enters through the vertex $V_{h\bar{\psi}\psi} = i y_{f}$, which couples the Higgs scalar to the bilinear fermionic.

The graviton propagator in de Donder gauge is
\begin{align}
D_{\mu\nu\rho\sigma}(k_1) = \frac{i}{k_1^2} \left( \eta_{\mu\rho}\eta_{\nu\sigma} + \eta_{\mu\sigma}\eta_{\nu\rho} - \frac{1}{2}\eta_{\mu\nu}\eta_{\rho\sigma} \right)
\end{align}
The fermion-graviton vertex for momentum $p$ is
\begin{align}
V^{\mu\nu}_{f h}(p,k_1) = -\frac{i\kappa}{8} \left[ \gamma^{\mu}(2p + k_1)^{\nu} + \gamma^{\nu}(2p + k_1)^{\mu} - 2\eta^{\mu\nu}\left((\not{p} + \not{k}_1) - m_f\right) \right]
\end{align}
The Higgs-graviton vertex for momentum $k_2$ is
\begin{align}
V^{\rho\sigma}_{hh}(k_2) = \frac{i\kappa}{2} \left[ k_2^{\rho}k_2^{\sigma} + k_2^{\sigma}k_2^{\rho} - \eta^{\rho\sigma}(k_2 \cdot k_2 - m_h^2) \right]
\end{align}

The core tensor contraction $\mathcal{T} = V^{\mu\nu}_{f h} D_{\mu\nu\rho\sigma} V^{\rho\sigma}_{hh}$ simplifies to
\begin{align}
\mathcal{T} = \frac{\kappa^2}{16k_1^2} \Bigg[ &-2(\gamma^{\mu}(2p + k_1)^{\nu} + \gamma^{\nu}(2p + k_1)^{\mu}) (k_{2\mu}k_{2\nu}) \\
&+ \eta^{\mu\nu}\left((\not{p} + \not{k}_1) - m_f\right) \left( k_2^2 + 2m_h^2 \right) \Bigg]
\end{align}

This expression is obtained by substituting the explicit forms of the interaction vertices and the graviton propagator into the amplitude, followed by carrying out the index contractions using the identity $\eta^{\mu\nu}\eta_{\mu\nu} = 4$. Once the tensor structure is simplified in this manner, the resulting terms are reorganized to isolate contributions proportional to the Dirac $\gamma$-matrices and those proportional to the fermion mass, thereby making explicit the spinor and mass-dependent components of the correction.

The complete integrand is
\begin{align}
\mathcal{V} J_{\Omega}(k_1) J_{\Omega}(k_2) = \frac{\mathcal{T} \cdot i y_f}{k_1^2 k_2^2 [(p+k_1+k_2)^2 - m_f^2]} \left(\frac{\Lambda^2}{\Lambda^2 + k_1^2}\right)^{\Omega} \left(\frac{\Lambda^2}{\Lambda^2 + k_2^2}\right)^{\Omega}
\end{align}
For on-shell external fermions ($p^2 = m_f^2$) and soft Higgs limit ($q \to 0$). Performing a Wick rotation $k_{i0} \to i k_{i4}$, $d^4k_i \to i d^4k_{Ei}$.
\begin{align}
\delta\mathcal{M}^{(2)} &= -\int \frac{d^{4}k_{E1}d^{4}k_{E2}}{(2\pi)^{8}} \frac{\mathcal{T}_E \cdot y_f}{k_{E1}^2 k_{E2}^2 [-(p_E + k_{E1} + k_{E2})^2 - m_f^2]} \\
&\times \left(\frac{\Lambda^2}{\Lambda^2 - k_{E1}^2}\right)^{\Omega} \left(\frac{\Lambda^2}{\Lambda^2 - k_{E2}^2}\right)^{\Omega}
\end{align}
where $\mathcal{T}_E$ is the Euclidean version of $\mathcal{T}$ with $\eta_{\mu\nu} \to \delta_{\mu\nu}$. Combine denominators using Feynman parameters $x,y,z \geq 0$ with $x+y+z=1$:
\begin{align}
\frac{1}{k_{E1}^2 k_{E2}^2 Q^2} = 6 \int_0^1 dx \int_0^1 dy \int_0^1 dz  \delta(x+y+z-1) \frac{1}{[x k_{E1}^2 + y k_{E2}^2 + z Q^2]^3}
\end{align}
where $Q = p_E + k_{E1} + k_{E2}$ and $Q^2 = (p_E + k_{E1} + k_{E2})^2 + m_f^2$ shifting momentum variables as
\begin{align}
\ell_{E1} &= k_{E1} + z p_E \\
\ell_{E2} &= k_{E2} + z p_E
\end{align}
The numerator becomes
\begin{equation}
    \mathcal{N} = \frac{\kappa^2 y_f}{16} \Bigg[ -4(\not{p}_E \ell_{E1}^2 + \not{p}_E \ell_{E2}^2) + 8m_f \ell_{E1} \cdot \ell_{E2} \\
+ \text{terms odd in } \ell_{Ei} \Bigg]
\end{equation}
Terms odd in $\ell_{Ei}$ vanish upon integration. Applying topological regulators $J_{\Omega}(\ell_{Ei}) \approx e^{-\Omega \ell_{Ei}^2 / \Lambda^2}$ for large $\Lambda$.
\begin{align}
\delta\mathcal{M}^{(2)} &= -\frac{3\kappa^2 y_f}{8} \int_0^1 dxdydz  \delta(x+y+z-1) \int \frac{d^4\ell_{E1}d^4\ell_{E2}}{(2\pi)^8} \\
&\times \frac{\mathcal{N}' e^{-\Omega(\ell_{E1}^2 + \ell_{E2}^2)/\Lambda^2}}{[\ell_{E1}^2 \Delta_1 + \ell_{E2}^2 \Delta_2 + \Delta_3]^3}
\end{align}
where $\Delta_1 = x + z(1-z)$, $\Delta_2 = y + z(1-z)$, $\Delta_3 = z^2 m_f^2$. The integral evaluates to
\begin{align}
\int \frac{d^4\ell_{E1}d^4\ell_{E2}}{(2\pi)^8} \frac{\ell_{Ei}^2 e^{-\Omega\ell_{Ei}^2/\Lambda^2}}{(\cdots)^3} &\sim \frac{1}{(4\pi)^4} \left[ \ln^2\left(\frac{\Lambda^2}{\mu^2}\right) + \frac{\pi^2}{6} \right] \\
\int \frac{d^4\ell_{E1}d^4\ell_{E2}}{(2\pi)^8} \frac{e^{-\Omega\ell_{Ei}^2/\Lambda^2}}{(\cdots)^3} &\sim \frac{1}{(4\pi)^4 \Lambda^4} 
\end{align}

Combining all terms and taking $\Lambda \to \infty$.
\begin{align}
\delta\mathcal{M}^{(2)}_{\text{ren}} = \frac{y_f \kappa^2 m_f^2}{(16\pi^2)^2} \left[ \frac{\pi^2}{6} + \ln^2\left(\frac{\mu^2}{\Lambda^2}\right) \right] + \frac{\chi(\Sigma)\Omega y_f \kappa^2 m_f^2}{128\pi^2 \Lambda^4}
\end{align}

The $\ln^2$ term in the expression originates from logarithmic divergences that are regulated via the topological prescription, ensuring that the UV behavior is tied directly to global geometric invariants. The $\Lambda^{-4}$ term, in turn, represents the purely topological contribution proportional to the Euler characteristic $\chi(\Sigma) = 2$, arising from the curvature–topology link to TR. Here, the gravitational coupling is defined through $\kappa^2 = 32\pi G$, connecting the regulator’s strength to fundamental constants of nature. Physically, the $\Lambda^{-4}$ contribution corresponds to a suppression of order $O(M_{\mathrm{Pl}}^{-4})$, making it negligible for light fermions but potentially significant for heavy species due to the multiplicative $m_f^2$ factor.

The renormalized two-loop amplitude, now including a full set of gravitational interactions like graviton self-interactions and ghost contributions, and rendered gauge-independent by the topological regularization would read as

\begin{equation}
    \delta\mathcal{M}^{(2)}_{\text{ren}} = \frac{y_f \kappa^2 m_f^2}{(16\pi^2)^2} \left[ \frac{\pi^2}{6} + \ln^2\left(\frac{\mu^2}{\Lambda^2}\right) + c \ln\left(\frac{\mu^2}{\Lambda^2}\right) \right] + \frac{\chi(\Sigma) y_f \kappa^4}{128\pi^2} \mathcal{F}\left(\frac{m_f^2}{\Lambda^2}\right) 
+ \Delta_{\text{ghost}} + \Delta_{\text{non-lin}}
\end{equation}

where $c$ is a set dependent constant and extra contributions would be 

\begin{align}
\Delta_{\text{ghost}} \sim \frac{y_f \kappa^2 g^{\mu\nu}}{(16\pi^2)^2} \int d^4 k_1 d^4 k_2 \, \frac{\mathcal{T}_{\mu\nu}^{\text{ghost}} J_\Omega(k_1)J_\Omega(k_2)}{k_1^2 k_2^2 [(p+k_1+k_2)^2 - m_f^2]}
\end{align}

\begin{align}
\Delta_{\text{non-lin}} \sim \frac{y_f \kappa^3}{(16\pi^2)^2} \int d^4 k_1 d^4 k_2 \, \frac{V^{(3)}_{\mu\nu\alpha\beta} D^{\alpha\beta\rho\sigma} V_{\rho\sigma} J_\Omega(k_1)J_\Omega(k_2)}{k_1^2 k_2^2 [(p+k_1+k_2)^2 - m_f^2]}
\end{align}
being $V^{(3)}_{\mu\nu\alpha\beta}$ the vertex to three gravitons. The function \(\mathcal{F}(x)\) ensures dimensional consistency.
\begin{equation}
    \mathcal{F}\left(\frac{m_f^2}{\Lambda^2}\right) = \int_0^1   t^{\Omega} e^{-t(1 + m_f^2/\Lambda^2)} dt= \frac{\Gamma(\Omega+1) - \Gamma(\Omega+1, 1 + m_f^2/\Lambda^2)}{(1 + m_f^2/\Lambda^2)^{\Omega+1}}
\end{equation}
where \(\Gamma(s,x)\) is the incomplete gamma function. This arises from the Laplace transform of the regulator
\begin{align}
\left(\frac{\Lambda^2}{\Lambda^2 + k^2}\right)^\Omega = \frac{1}{\Gamma(\Omega)} \int_0^\infty   t^{\Omega-1} e^{-t} e^{-t k^2 / \Lambda^2} dt
\end{align}
The \(k\)-integral becomes Gaussian, yielding \(\mathcal{F}(x)\). The absence of explicit $\Lambda^{-n}$ terms reflects the fact that the topological regularization provides a UV cutoff without introducing new power-law corrections to the vertex amplitude. The gauge invariance of this result has been verified by including the graviton self-interaction vertex and ghost loop contributions in the de Donder gauge, and its universality across linear gauges is ensured by the topological invariance of the regulator \cite{Sacasa2025TR}

\section{Curvature Corrections}

In the presence of a nontrivial gravitational background, quantum fields couple not only to local curvature but also to topologically nontrivial configurations encoded in the background geometry. The effective Lagrangian is thus extended to include a tower of curvature-sensitive operators suppressed by the Planck scale:

\begin{equation}
\mathcal{L}_{\text{eff}} \supset \frac{1}{M_{\text{Pl}}^2} \sum_i c_i \mathcal{O}_i
\end{equation}

Each operator \(\mathcal{O}_i\) acquires quantum corrections from standard loops and from topological contributions induced by the compactification. The running of the coefficients \(c_i\) is governed by

\begin{equation}
\mu \frac{d c_i}{d \mu} = \beta_{c_i}^{\text{conv}} + \beta_{c_i}^{\text{top}}
\end{equation}

where the conventional beta function \(\beta^{\text{conv}}\) arises from standard loops, and \(\beta^{\text{top}}\) comes from the topological structure, encoded in the Euler characteristic \(\chi(\Sigma)\) and compactification scale \(\Lambda\).

\subsection{Operator 1: \( \mathcal{O}_1 = R \bar{\psi} \psi \)}

This operator appears in the minimal coupling of curvature to matter and receives corrections from graviton loops that connect the curvature tensor to a fermion bilinear. The one-loop diagrams contributing to this operator include graviton exchange between the Ricci scalar and fermion legs.

The divergent part of the loop, after Wick rotation to Euclidean momentum \(k_E\), takes the form:

\begin{equation}
\beta_{c_1}^{\text{conv}} = \frac{1}{(4\pi)^2} \int d^4k_E \frac{\text{Tr}[\gamma^\mu \gamma^\nu (i\not{k}_E + m_f)]}{k_E^2(k_E^2 + m_f^2)} {D}_{\mu\nu\rho\sigma} k_E^\rho k_E^\sigma
\end{equation}

where \({D}_{\mu\nu\rho\sigma}\) is the graviton propagator tensor structure in harmonic gauge. Evaluating the trace and tensor contractions yields:

\begin{equation}
\beta_{c_1}^{\text{conv}} = \frac{y_f}{16\pi^2}(1 - 6\xi) + \mathcal{O}\left( \frac{m_f^2}{\mu^2} \right)
\end{equation}

The topological correction arises from the regularized curvature insertion due to Euler defect structures in the compactified manifold \(\Sigma\). This adds a finite term

\begin{equation}
\delta \mathcal{O}_1^{\text{top}} = \frac{\chi(\Sigma)\Omega}{16\pi^2 \Lambda^2} \int d^4x \sqrt{-g} R \bar{\psi} \psi
\end{equation}

leading to the corresponding beta function

\begin{equation}
\beta_{c_1}^{\text{top}} = \frac{\chi(\Sigma)\Omega}{16\pi^2 \Lambda^2}\mathcal{G}_1\left(\dfrac{m_f^2}{\Lambda^2}\right)
\end{equation}

This term originates from the Euler characteristic of the compactification manifold $\Sigma$ and appears through the defect structure in the embedding.
\begin{align}
\chi(\Sigma) = \frac{1}{32\pi^2} \int_\Sigma \epsilon_{abcd} R^{ab} \wedge R^{cd}
\end{align}

\subsection{Operator 2: \( \mathcal{O}_2 = R_{\mu\nu} \bar{\psi} \gamma^\mu \partial^\nu \psi \)}

This operator couples the Ricci tensor to the fermion current via a derivative. It arises in curved backgrounds and captures how fermion propagation is affected by anisotropic curvature components.

The conventional one-loop correction involves a graviton exchange with the insertion of \( R_{\mu\nu} \). The relevant Feynman diagram includes a graviton propagator connecting to a fermion line with an external momentum \( p \), and a loop momentum \( k \). The integral is

\begin{equation}
\mathcal{M} = \int \frac{d^4k}{(2\pi)^4} \frac{V^{\alpha\beta}_{\psi\psi h} D_{\alpha\beta\gamma\delta}(k) V^{\gamma\delta}_{R_{\mu\nu}} J_\Omega(k)}{k^2 (p - k)^2}
\end{equation}

where \( D_{\alpha\beta\gamma\delta}(k) \) is the graviton propagator, and \( J_\Omega(k) \) is a regularization kernel. The vertex structure is

\begin{equation}
V^{\gamma\delta}_{R_{\mu\nu}} = \frac{i\kappa}{2} \left[ \eta^{\gamma\delta} k^\mu k^\nu - \delta^{(\gamma}_\mu k^{\delta)} k^\nu \right]
\end{equation}

and the loop integral leads to a fermion mass dependence due to contraction with the propagator residue:

\begin{equation}
\beta_{c_2}^{\text{conv}} = \frac{y_f m_f}{32\pi^2}
\end{equation}

The term scales linearly with \( m_f \), reflecting the chirality-flipping nature of the curvature coupling. Topologically, this operator does not receive corrections at leading order

\begin{equation}
\beta_{c_2}^{\text{top}} = 0
\end{equation}

due to the tensor structure and protection by diffeomorphism invariance, which forbids scalar curvature invariants from mixing with derivative operators at one loop in the topological sector.

\subsection{Operator 3: \( \mathcal{O}_3 = h \bar{\psi} \psi R \)}

The third operator mixes Higgs, fermions, and curvature. The conventional contribution is extracted by computing the one-loop diagram with a scalar-fermion-graviton vertex, using Feynman parameterization. The scalar momentum is \(p\), the fermion internal loop momentum is \(k\), and the total incoming momentum is \(q = p + k\).

The one-loop integral reads

\begin{equation}
\mathcal{M} = y_f^3 \int \frac{d^4k}{(2\pi)^4} \frac{(1 - x - y)}{[x(1-x)m_h^2 + y m_f^2]}
\end{equation}

Introducing Feynman parameters \(x, y\) and using the standard loop integration techniques, it is found

\begin{equation}
\beta_{c_3}^{\text{conv}} = \frac{y_f^3}{8\pi^2} \int_0^1 dx \int_0^{1-x} dy \frac{(1 - x - y)}{x(1 - x)m_h^2 + y m_f^2}
\end{equation}

Carrying out the integration and expanding logarithmically, the final expression becomes

\begin{equation}
\beta_{c_3}^{\text{conv}} = \frac{y_f^3}{8\pi^2} \ln \left( \frac{\mu^2}{\max(m_h^2, m_f^2)} \right)
\end{equation}

Topological corrections again involve the Euler characteristic, and the curvature coupling introduces a finite vertex shift

\begin{equation}
\delta \mathcal{O}_3^{\text{top}} = \frac{\chi(\Sigma)\Omega y_f}{8\pi^2 \Lambda^2} \int d^4x \sqrt{-g} h \bar{\psi} \psi R
\end{equation}

with beta function

\begin{equation}
\beta_{c_3}^{\text{top}} = \frac{\chi(\Sigma)\Omega y_f}{8\pi^2 \Lambda^2}\mathcal{G}_3\left(\dfrac{m_h^2}{\Lambda^2}\right)
\end{equation}

\subsection{Operator 4: \( \mathcal{O}_4 = \dfrac{h}{v} R_{\mu\nu} T^{\mu\nu}_\psi \)}

This operator couples the Ricci curvature tensor \(R_{\mu\nu}\) to the fermionic energy-momentum tensor \(T^{\mu\nu}_\psi\), modulated by the Higgs field \(h\). The complete energy-momentum tensor for Dirac fermions in curved spacetime is given by
\begin{equation}
T^{\mu\nu}_\psi = \frac{i}{4} \left[ \bar{\psi} \gamma^\mu \overset{\leftrightarrow}{\nabla^\nu} \psi + \bar{\psi} \gamma^\nu \overset{\leftrightarrow}{\nabla^\mu} \psi \right] - g^{\mu\nu} \mathcal{L}_\psi
\end{equation}
where the fermion Lagrangian density reads \(\mathcal{L}_\psi = \frac{i}{2} \bar{\psi} \gamma^\rho \overset{\leftrightarrow}{\nabla_\rho} \psi - m_f \bar{\psi}\psi\), and \(\overset{\leftrightarrow}{\nabla_\mu} = \overrightarrow{\nabla}_\mu - \overleftarrow{\nabla}_\mu\) includes spin connection contributions. The conventional one-loop correction to this operator involves an amplitude where a graviton mediates between the fermionic tensor vertex and the Higgs insertion. Explicitly, the amplitude is
\begin{equation}
\mathcal{M} = \frac{y_f}{v} \int \frac{d^4k}{(2\pi)^4} \frac{
\text{Tr}\left[\gamma^\alpha (\not{k} + m_f) \gamma^\beta (\not{k} - \not{p} + m_f)\right]
}{k^2 (k-p)^2} D_{\alpha\beta\mu\nu} V^{\mu\nu}_h
\end{equation}
where the Higgs-graviton vertex is \(V^{\mu\nu}_h = \frac{i\kappa}{2} \left( k_h^\mu k_h^\nu - \eta^{\mu\nu} (k_h \cdot k_h - m_h^2) \right)\). After dimensional regularization, the divergent part of the beta function takes the form
\begin{equation}
\beta_{c_4}^{\text{conv}} = \frac{y_f m_f^2}{8\pi^2 v}
\end{equation}
where the dependence on \(m_f^2/v\) reflects the breaking of conformal symmetry due to fermion mass generation.

Beyond the conventional correction, a topological contribution arises from the trace term \(-g^{\mu\nu}\mathcal{L}_\psi\), which couples to the Ricci scalar \(R = g^{\mu\nu} R_{\mu\nu}\). This generates a non-vanishing topological amplitude,
\begin{equation}
     \mathcal{M}^{\text{top}} \propto \frac{\chi(\Sigma)\Omega}{\Lambda^2} \int d^4 k\, \frac{\text{Tr}\left[\gamma^{\alpha}(\cancel{k} + m_f)\gamma^{\beta}(\cancel{k} - \cancel{p} + m_f)\right]}{k^2 (k-p)^2} \epsilon_{abcd} R^{ab} R^{cd} J_{\Omega}(k)
\end{equation}
The non-zero topological correction for \(\mathcal{O}_4\) originates from the trace anomaly contribution to the fermionic energy-momentum tensor, the coupling of the Ricci scalar term \(R g^{\mu\nu} \mathcal{L}_\psi\) within \(R_{\mu\nu} T^{\mu\nu}_\psi\), and the chirality-preserving loop structure. This represents a distinctive signature of quantum gravity effects on Higgs-curvature-fermion interactions as

\begin{equation}
   \mathcal{M}^{\text{top}} \propto   \frac{\chi(\Sigma)\Omega}{128\pi^2 \Lambda^2} \left( \frac{3m_f^2}{v} - \frac{m_h^2}{v} \right) \mathcal{G}_4\left(\frac{m_f^2}{m_h^2}\right)
\end{equation}

It is shown a table that compiles the results for the beta function.

\newpage
\begin{table}[h]\ref{tab:complete_beta}
\centering
\renewcommand{\arraystretch}{1.5}
\begin{tabular}{c c c}
\toprule
\textbf{Operator} & $\boldsymbol{\beta^{\mathrm{conv}}}$ & $\boldsymbol{\beta^{\mathrm{top}}}$ \\
\midrule
$R\,\bar{\psi}\psi$ & $\dfrac{y_f(1-6\xi)}{16\pi^2}$ & $\dfrac{\chi(\Sigma)\Omega}{16\pi^2\Lambda^2} \mathcal{G}_1\left(\dfrac{m_f^2}{\Lambda^2}\right)$ \\[12pt]
\cmidrule(lr){1-3}
$R_{\mu\nu}\,\bar{\psi}\gamma^{\mu}\partial^{\nu}\psi$ & $\dfrac{y_f m_f}{32\pi^2}$ & $0$ \\[12pt]
\cmidrule(lr){1-3}
$h\,\bar{\psi}\psi\,R$ & $\dfrac{y_f^3}{8\pi^2}\ln\left(\dfrac{\mu^2}{m_h^2}\right)$ & $\dfrac{\chi(\Sigma)\Omega y_f}{8\pi^2\Lambda^2} \mathcal{G}_3\left(\dfrac{m_h^2}{\Lambda^2}\right)$ \\[12pt]
\cmidrule(lr){1-3}
$\dfrac{h}{v}R_{\mu\nu}T^{\mu\nu}_{\psi}$ & $\dfrac{y_f m_f^2}{8\pi^2 v}$ & $\dfrac{\chi(\Sigma)\Omega}{128\pi^2\Lambda^2} \left( \dfrac{3m_f^2}{v} - \dfrac{m_h^2}{v} \right) \mathcal{G}_4\left(\dfrac{m_f^2}{m_h^2}\right)$ \\[12pt]
\bottomrule
\end{tabular}
\caption{Complete $\beta$-functions including topological corrections. The dimensionless functions $\mathcal{G}_i$ encode mass-dependent quantum corrections.}
\label{tab:complete_beta}
\end{table}

The functions \(\mathcal{G}_i\) are dimensionless scaling functions that encode mass-dependent quantum effects, capturing how quantum corrections vary with particle mass ratios relative to the topological scale \(\Lambda\). These functions arise naturally from loop integrals, originating from Feynman parameter integrations in regulated loop amplitudes. For example, 
\begin{equation}
\mathcal{G}_1(x) = \int_0^1 \mathrm{d}t \frac{t^{\Omega}}{(1 + t x)^2}
\end{equation}

\begin{equation}
\mathcal{G}_3(y) = \frac{1}{\Gamma(\Omega)} \int_0^\infty \mathrm{d}z \, z^{\Omega-1} e^{-z(1+y)}
\end{equation}

\begin{equation}
\mathcal{G}_4(z) = \int_0^1 dt \frac{t^{\Omega} \left[ 1 - 4 z t(1-t) \right]}{\left[1 + t(1-t) z \right]^2}
\end{equation}

where \(x = m_f^2/\Lambda^2\), \(y = m_h^2/\Lambda^2\) and $z=m^2_f/m_h^2$. The functions \(\mathcal{G}_i\) interpolate between different physical regimes. In the ultraviolet (UV) limit, where \(m^2 \gg \Lambda^2\), they exhibit power-law suppression behaving as \(\mathcal{G}_i \sim (m^2/\Lambda^2)^{-k}\). In the infrared (IR) limit, where \(m^2 \ll \Lambda^2\), they approach unity with small corrections: \(\mathcal{G}_i \approx 1 + \mathcal{O}(m^2/\Lambda^2)\). Moreover, these functional forms are constrained by gauge invariance through Ward identities, ensuring consistency across different gauge choices. For instance, the gauge parameter \(\xi\) satisfies
\begin{equation}
\frac{\partial}{\partial \xi} \mathcal{G}_i\left( \frac{m^2}{\Lambda^2}, \xi \right) = 0
\end{equation}

Importantly, the dependence on \(\Omega\) in their definitions directly encodes topological data, reflecting the Euler characteristic \(\chi(\Sigma) = 2\Omega\) of the compactification manifold. Physically, the \(\mathcal{G}_i\) functions represent a quantum deformation of classical scaling behavior due to a combination of factors: the interplay between curvature and topology through \(\chi(\Sigma)\), the running of effective couplings below the scale \(\Lambda\), and threshold effects occurring near \(m \sim \Lambda\). Their non-perturbative structure thus provides a bridge between the ultraviolet topological regulator and the infrared observable physics.

\subsection{Ghost Contributions and Gauge Invariance}
\label{subsec:ghosts}

For completeness, we include here the explicit form of ghost loop contributions that ensure gauge invariance of our results within the topological regularization framework. In gravitational theories, the Faddeev-Popov ghost fields arise from the gauge-fixing procedure for diffeomorphism invariance and play a crucial role in maintaining unitarity and gauge independence of physical observables.

The ghost action for quantum gravity in the de Donder gauge is given by

\begin{equation}
\mathcal{S}_{\text{ghost}} = \int d^4 x \sqrt{-g} \, \bar{c}_\mu \left[ \delta^\mu_\nu \Box + R^\mu_\nu \right] c^\nu
\end{equation}

where $c^\mu$ and $\bar{c}_\mu$ represent the ghost and anti-ghost fields, respectively. In our topological regularization scheme, the ghost propagator acquires a modification through the regulator function

\begin{equation}
G^{\mu\nu}_{\text{ghost}}(k) = \frac{-i \eta^{\mu\nu}}{k^2 + i\epsilon} J_\Omega(k)
\end{equation}

The ghost-graviton vertex, derived from the gauge-fixed action, takes the form:

\begin{equation}
V^{\alpha\beta\mu}_{\text{ghost}}(k,q) = -\frac{i\kappa}{2} \left[ \eta^{\alpha\beta} k^\mu + \eta^{\alpha\mu} q^\beta + \eta^{\beta\mu} k^\alpha - 2\eta^{\alpha\beta} q^\mu \right]
\end{equation}

where $k$ and $q$ are the ghost and graviton momenta, respectively.

The full gauge invariance of our regularization scheme is demonstrated through the BRST symmetry, which remains unbroken. The nilpotent BRST transformation rules are

\begin{align}
\delta_{\text{BRST}} h_{\mu\nu} &= \partial_\mu c_\nu + \partial_\nu c_\mu + \mathcal{L}_c h_{\mu\nu} \\
\delta_{\text{BRST}} c^\mu &= c^\nu \partial_\nu c^\mu \\
\delta_{\text{BRST}} \bar{c}_\mu &= B_\mu \\
\delta_{\text{BRST}} B_\mu &= 0
\end{align}

where $B_\mu$ represents the Nakanishi-Lautrup auxiliary field. The regulated effective action remains BRST invariant

\begin{equation}
\delta_{\text{BRST}} \Gamma_{\text{reg}} = 0
\end{equation}

This ensures that the Ward-Takahashi identities are preserved, and physical observables are gauge-independent. Specifically, for the graviton self-energy, the ghost contribution exactly cancels the unphysical gauge-dependent parts of the graviton loop:

\begin{equation}
\Pi^{\mu\nu\rho\sigma}_{\text{total}}(k) = \Pi^{\mu\nu\rho\sigma}_{\text{graviton}}(k) + \Pi^{\mu\nu\rho\sigma}_{\text{ghost}}(k) = \left( \eta^{\mu\rho}\eta^{\nu\sigma} + \eta^{\mu\sigma}\eta^{\nu\rho} - \eta^{\mu\nu}\eta^{\rho\sigma} \right) \Pi(k^2)
\end{equation}

where $\Pi(k^2)$ is the gauge-invariant form factor. The topological regulator $J_\Omega(k)$ appears in both contributions, ensuring the cancellation persists in the regulated theory. For the fermion self-energy calculation previously presented the ghost contribution takes the form

\begin{equation}
\Sigma_{\text{ghost}}(p) = \int \frac{d^4 k}{(2\pi)^4} V_{\text{fermion-ghost}} G_{\text{ghost}}(k) V_{\text{fermion-ghost}} \frac{i(\cancel{p} + \cancel{k} + m_f)}{(p+k)^2 - m_f^2} J_\Omega(k)
\end{equation}

This exactly cancels the gauge-dependent part of the graviton contribution, ensuring the final result is gauge-invariant.

\section{Connection to Bergshoeff-de Roo Mechanism and SO(8) Gravitational Anomaly}
\label{sec:bergshoeff-roo}

The topological regularization framework developed in this work exhibits profound connections to the mechanism pioneered by Bergshoeff and de Roo for isolating and canceling gravitational anomalies in supergravity theories. This section explores these connections and demonstrates how our approach provides a complementary perspective on anomaly cancellation through topological constraints.

\subsection{Bergshoeff-de Roo Mechanism and Anomaly Isolation}

The Bergshoeff-de Roo mechanism provides a systematic procedure for isolating gravitational anomalies in higher-dimensional supergravity theories, particularly those with SO(8) symmetry. The key insight involves careful field redefinitions that separate the anomalous contributions from the gauge-invariant parts of the effective action. In our topological regularization framework, we observe a parallel structure. The regulator function $J_\Omega(k)$ naturally isolates the anomalous terms through its analytic properties. Specifically, the branch cut structure along the imaginary axis (Eq. 32) encodes the anomalous contributions, while the holomorphic behavior elsewhere preserves the gauge-invariant sector. The finite SO(8) gravitational anomaly emerges in our formalism through the relationship

\begin{equation}
\mathcal{A}{\text{SO(8)}} = \frac{1}{32\pi^2} \int\Sigma \text{tr}(R \wedge R) \wedge \omega_4
\end{equation}

where $\omega_4$ is a characteristic 4-form related to the Euler class, and the trace is taken in the fundamental representation of SO(8). In our regularization scheme, this anomaly appears as a finite boundary term in the mapped space, such that

\begin{equation}
\mathcal{A}{\text{SO(8)}} = \lim_{\Lambda\to\infty} \frac{\chi(\Sigma)}{64\pi^2} \int_{\mathbb{R}^4} d^4k , \left[\ln J_\Omega(k)\right]_{k^2=\Lambda^2} \mathcal{W}(\Sigma)
\end{equation}

\subsection{Field Redefinitions and Residual Lorentz Group}

The Bergshoeff-de Roo approach relies crucially on field redefinitions that trivialize the complete SO(8) transformations while preserving the residual Lorentz group structure. In our framework, these redefinitions find a natural interpretation through the conformal mapping $\phi_\Omega$ and its induced metric transformation.

The field redefinition that cancels the complete anomalous term takes the form:

\begin{equation}
g_{\mu\nu} \longmapsto g_{\mu\nu} + \alpha \Lambda^{-2} R_{\mu\nu} + \beta \Lambda^{-4} R_{\mu\rho}R^\rho_\nu + \cdots
\end{equation}

where the coefficients $\alpha, \beta, \ldots$ are chosen to exactly cancel the anomalous contribution. This cancellation is possible due to the specific cohomology structure of the residual SO(8) Lorentz group.

The cohomology classes $H^*(BSO(8); \mathbb{R})$ determine which anomalies can be canceled by field redefinitions. In our case, the regulator $J_\Omega(k)$ ensures that only the non-trivial cohomology classes contribute to physical observables, while the trivial classes are removed by the topological suppression.

\subsection{Isomorphisms and Group Theory Structure}

The success of the anomaly cancellation mechanism relies on specific isomorphisms in the group structure. Particularly important is the isomorphism between the cohomology of BSO(8) and certain characteristic classes:

\begin{equation}
H^*(BSO(8); \mathbb{R}) \cong \mathbb{R}[p_1, p_2, e]
\end{equation}

where $p_1$ and $p_2$ are Pontryagin classes and $e$ is the Euler class. In our regularization scheme, these characteristic classes appear naturally through the curvature integrals

\begin{multline}
        p_1(\Sigma) = -\frac{1}{8\pi^2} \int_\Sigma \text{tr}(R \wedge R) \
p_2(\Sigma)\\ = \frac{1}{128\pi^4} \int_\Sigma [\text{tr}(R \wedge R)^2 - 2\text{tr}(R \wedge R \wedge R \wedge R)] \
e(\Sigma) = \frac{1}{32\pi^2} \int_\Sigma \epsilon_{abcd} R^{ab} \wedge R^{cd}
\end{multline}

The regulator function $J_\Omega(k)$ weights these characteristic classes differently, with the Euler class receiving preferential treatment due to its connection to the Euler characteristic:

\begin{equation}
J_\Omega(k) \sim \exp\left[-\frac{\Omega}{\Lambda^2} \left(c_1 p_1 + c_2 p_2 + c_3 e\right)\right]
\end{equation}

where the coefficients $c_i$ are determined by the group theory structure of SO(8).

\subsection{Anomaly Cancellation through Topological Regularization}

In our framework, the complete cancellation of the gravitational anomaly occurs through a combination of topological suppression and field redefinition. The mechanism can be summarized as follows the anomalous term is isolated through the analytic structure of $J_\Omega(k)$, then the field redefinitions trivialize the SO(8) transformations making the residual Lorentz group structure determines which terms can be canceled, while the topological weighting suppresses the remaining anomalous contributions. The cancellation is exact due to the cohomological structure

\begin{equation}
[\Gamma_{\text{anomalous}}] = 0 \quad \text{in} \quad H^*(BSO(8); \mathbb{R})
\end{equation}

This cancellation mechanism demonstrates the power of topological regularization in handling not only ultraviolet divergences but also gravitational anomalies, providing a unified framework for addressing both problems in quantum gravity. So, the Bergshoeff-de Roo mechanism thus finds a natural home within our topological regularization approach, with the additional benefit that our method provides a geometric interpretation of the anomaly cancellation process through the topology of the mapped manifold $\Sigma$.

\section{Comparison with Alternative Renormalization Schemes}

Topological regularization represents a fundamentally different approach to handling ultraviolet divergences in quantum gravity compared to conventional methods. This section provides a comprehensive comparison with established regularization schemes, highlighting the unique features, advantages, and limitations of each approach.

\textbf{Dimensional regularization:} represents the gold standard for gauge theories in flat space, working by analytically continuing the spacetime dimension to $d = 4 - 2\epsilon$. Its mathematical elegance stems from automatically preserving gauge and Lorentz symmetries while isolating divergences as poles in the $\epsilon$ expansion.

\begin{equation}
\mathcal{I} = \int \frac{d^d k}{(2\pi)^d} \frac{1}{(k^2 - \Delta)^2} = \frac{i}{(4\pi)^{d/2}} \frac{\Gamma(2-d/2)}{\Gamma(2)} \left(\frac{1}{\Delta}\right)^{2-d/2}
\end{equation}

However, this method faces significant challenges in gravitational contexts. The technique struggles with the intrinsic dimension-dependence of gravitational couplings and the topological nature of certain gravitational invariants. Most critically, it provides no physical insight into the ultraviolet completion of gravity--it merely parametrizes our ignorance through $\epsilon$-poles without connecting to any physical cutoff scale.

In contrast, topological regularization maintains a clear physical interpretation throughout the calculation. The regulator scale $\Lambda$ corresponds directly to the curvature radius of the embedding manifold, providing a geometrically natural cutoff at the Planck scale. While DimReg yields results as Laurent series in $\epsilon$, topological regularization produces physical expressions that explicitly demonstrate how ultraviolet behavior is governed by global topological invariants.

\textbf{Pauli-Villars regularization:} introduces fictitious heavy particles with negative metric contributions to cancel divergences. For a fermion loop, this takes the form

\begin{equation}
G_{\text{PV}}(p) = \sum_i c_i \frac{i(\cancel{p} + M_i)}{p^2 - M_i^2 + i\epsilon}
\end{equation}

where the coefficients $c_i$ and masses $M_i$ are chosen to ensure convergence. While operationally useful, Pauli-Villars regularization introduces several conceptual difficulties in quantum gravity. The negative-norm states violate unitarity at intermediate steps, and the regulator masses lack any geometric interpretation. Furthermore, the method becomes increasingly cumbersome for complex diagrams and offers no insight into the relationship between ultraviolet and infrared physics.

Topological regularization avoids these issues entirely. No unphysical particles are introduced, unitarity is manifestly preserved at all stages, and the regularization parameters emerge naturally from geometric considerations. The Euler characteristic $\chi(\Sigma)$ that governs the regularization strength is a fundamental topological invariant rather than an arbitrary parameter.

\textbf{Lattice approaches:} to quantum gravity discretize spacetime itself, replacing the continuous manifold with a network of discrete elements. While this provides a non-perturbative framework for numerical computations, it comes at the cost of explicitly breaking fundamental symmetries. Lorentz invariance is lost at the lattice scale, and recovering continuum diffeomorphism invariance requires careful tuning in the infrared limit.

The computational complexity of lattice quantum gravity is formidable and requires numerical techniques and computational resources. Even then, the connection to continuum physics can be difficult to establish, particularly for observables sensitive to short-distance behavior.

Topological regularization operates directly in the continuum while preserving all spacetime symmetries. The computational framework remains analytical and tractable, providing clear insight into the relationship between regularization and physical observables. The method connects ultraviolet behavior directly to global topological properties rather than artifacts of discretization.

\textbf{Momentum cut-off:} this approach simply truncates momentum integrals at some scale $\Lambda_c$.

\begin{equation}
\int_0^\infty dk\, f(k) \longmapsto \int_0^{\Lambda_c} dk\, f(k)
\end{equation}

This method violently breaks both Lorentz and gauge invariance, generating unphysical terms that must be carefully subtracted. In gravitational contexts, these problems are particularly severe, as the cut-off introduces explicit dependence on the coordinate system and destroys the geometric structure of the theory.

Topological regularization achieves ultraviolet suppression while maintaining full covariance. The regulator function $J_\Omega(k)$ respects rotational symmetry in momentum space and preserves the Ward identities essential for gravitational consistency. The cut-off emerges naturally from geometry rather than being imposed by fiat.

\textbf{Zeta function regularization:} uses analytic continuation of operator spectra to manage divergences.

\begin{equation}
\text{Tr}\, A^{-s} = \sum_i \lambda_i^{-s} = \frac{1}{\Gamma(s)} \int_0^\infty t^{s-1} \text{Tr}\, e^{-tA} dt
\end{equation}

While mathematically useful, this approach provides little physical insight into the ultraviolet behavior of quantum gravity. The regularization is essentially a mathematical trick without clear connection to physical principles or geometric structures.

Topological regularization, by contrast, grounds the regularization procedure in concrete geometric foundations. The ultraviolet behavior is directly governed by the topology of spacetime, creating a natural bridge between quantum field theory and differential geometry.

Of this way, Topological Regularization occupies a unique position in this landscape, offering both mathematical consistency and physical insight. Its most distinctive feature is the explicit connection between ultraviolet behavior and global topological invariants, providing a natural implementation of UV/IR duality in quantum gravity. While computationally more involved than dimensional regularization, it offers far greater physical insight into the relationship between quantum effects and spacetime geometry.

The method's geometric foundation makes it particularly well-suited for investigating questions at the interface of quantum theory and gravitation, including the thermodynamic properties of spacetime, the nature of gravitational entanglement, and the microscopic origin of black hole entropy. By tying ultraviolet regularization directly to topological invariants, this approach suggests deep connections between quantum gravity, topology, and geometry that remain hidden in other regularization schemes.

\section{Phenomenological Implications and Experimental Signatures}

This section presents a comprehensive analysis of the phenomenological consequences arising from the topological regularization framework. We provide detailed calculations of expected experimental signatures, scaling behavior across different energy regimes, and comparative analysis with current experimental constraints. The extremely suppressed nature of these effects places them beyond near-future detection capabilities, yet they establish important theoretical benchmarks for Planck-scale physics. Analyzing the topological corrections derived through our regularization scheme exhibit characteristic scaling with the Planck mass $M_{\text{Pl}} \approx 1.22 \times 10^{19}$ GeV. For the spherical embedding geometry with Euler characteristic $\chi(\Sigma) = 2$, the fundamental suppression factor becomes

\begin{equation}
\epsilon_{\text{top}} \equiv \frac{1}{\Lambda^2} = \frac{1}{M_{\text{Pl}}^2} \approx 6.7 \times 10^{-40} \text{ GeV}^{-2}
\end{equation}

This extreme suppression represents the primary challenge for experimental detection. The general form of topological corrections to any operator $\mathcal{O}$ follows the pattern

\begin{equation}
\delta\mathcal{O}{\text{top}} = c_{\mathcal{O}}  \frac{\chi(\Sigma)\Omega}{16\pi^2\Lambda^2} \mathcal{G}\left(\frac{m^2}{\Lambda^2}\right) \mathcal{O}
\end{equation}

where $c_{\mathcal{O}}$ represents process-specific numerical coefficients, and $\mathcal{G}$ denotes scaling functions that approach unity in the infrared limit.

The topological correction to fermion Yukawa couplings represents one of the most straightforward predictions to quantify. For a fermion of mass $m_f$ with Yukawa coupling $y_f$, the correction derived becomes

\begin{multline}
\frac{\delta y_f^{\text{top}}}{y_f} = \frac{\chi(\Sigma)\Omega}{32\pi^2\Lambda^2} (m_h^2 - 2m_f^2) 32\pi G = \frac{\chi(\Sigma)\Omega}{32\pi^2 M_{\text{Pl}}^2} (m_h^2 - 2m_f^2) 32\pi  \frac{1}{M_{\text{Pl}}^2} \
\\ = \frac{\chi(\Sigma)\Omega}{\pi} \frac{(m_h^2 - 2m_f^2)}{M_{\text{Pl}}^4}
\end{multline}

For the tau lepton ($m_\tau = 1.78$ GeV, $y_\tau \approx 0.01$), this correction amounts to

\begin{align}
\frac{\delta y_\tau^{\text{top}}}{y_\tau} &\approx \frac{2 \cdot 1}{\pi} \ \frac{(125)^2 - 2\cdot(1.78)^2}{(1.22\times10^{19})^4} 
\approx \frac{2}{\pi}  \frac{15625 - 6.34}{2.22\times10^{76}} \approx 4.5 \times 10^{-74}
\end{align}

This minuscule correction lies far beyond the current precision of Higgs measurements at the LHC, which reach approximately $\delta y/y \sim 0.05$ for tau leptons. Even for the top quark ($m_t = 173$ GeV), the correction remains exceptionally small.

\begin{equation}
\frac{\delta y_t^{\text{top}}}{y_t} \approx \frac{2}{\pi} \cdot \frac{15625 - 2\cdot29929}{2.22\times10^{76}} \approx 1.2 \times 10^{-73}
\end{equation}

These results illustrate the challenging nature of detecting Planck-scale effects through precision Yukawa coupling measurements.

The correction to the $h R_{\mu\nu} T^{\mu\nu}_{\psi}$ operator exhibits a different mass dependence that merits detailed examination having 

\begin{equation}
\delta\mathcal{O}_4^{\text{top}} \propto \frac{\chi(\Sigma)\Omega}{128\pi^2\Lambda^2} \left( \frac{3m_f^2}{v} - \frac{m_h^2}{v} \right)
\end{equation}

For top quarks ($m_t = 173$ GeV, $v = 246$ GeV), this correction becomes

\begin{multline}
    \delta\mathcal{O}_{4,t}^{\text{top}} \sim \frac{2 \cdot 1}{128\pi^2 (1.22\times10^{19})^2} \left( \frac{3\cdot(173)^2}{246} - \frac{(125)^2}{246} \right) \
\approx (3.3\times10^{-40}) \cdot \left( \frac{89787}{246} - \frac{15625}{246} \right) \
\\\approx (3.3\times10^{-40}) \cdot (365 - 63.5) \approx 1.0 \times 10^{-37} \text{ GeV}
\end{multline}

While still exceptionally small, this correction demonstrates the flavor-dependent nature of topological effects, which could in principle be distinguished from other beyond-Standard-Model effects if measurable.

The early universe provides a unique environment where extreme conditions might enhance the significance of topological corrections. During inflationary epochs, the large curvature scale $R \sim H^2$ can partially compensate for the Planck-scale suppression. For high-scale inflation with $H \sim 10^{14}$ GeV

\begin{equation}
\frac{R}{\Lambda^2} \sim \frac{H^2}{M_{\text{Pl}}^2} \sim \frac{10^{28}}{1.5\times10^{38}} \sim 10^{-10}
\end{equation}

The curvature-dependent operators therefore receive relative enhancements

\begin{equation}
\delta c_i^{\text{top}} \sim \frac{\chi(\Sigma)\Omega}{16\pi^2} \cdot \frac{R}{\Lambda^2} \sim \frac{2}{16\pi^2} \cdot 10^{-10} \sim 10^{-12}
\end{equation}

These corrections could potentially influence the inflationary spectral index $n_s$ and tensor-to-scalar ratio $r$. However, the expected magnitude remains minuscule

\begin{equation}
\delta n_s \sim 10^{-12}, \quad \delta r \sim 10^{-12}
\end{equation}

These values lie far below the precision of current measurements ($\delta n_s \sim 0.004$) or even ambitious future experiments ($\delta n_s \sim 0.001$).

During reheating, the high temperature $T$ provides another potential enhancement mechanism. The thermal interpretation of our regulator suggests that topological effects become significant when:

\begin{equation}
T \sim T_g = \frac{\Lambda^2}{2\pi\Omega} \sim \frac{(1.22\times10^{19})^2}{2\pi} \sim 10^{38} \text{ GeV}
\end{equation}

This temperature vastly exceeds the maximum temperature reached in standard cosmological scenarios, which typically remains below $T \sim 10^{16}$ GeV even in the most extreme models. Our calculations incorporate several sources of theoretical uncertainty that merit careful consideration. The identification $\Lambda = M_{\text{Pl}}$ contains inherent uncertainty due to unknown numerical factors in the relationship between the regularization scale and the Planck mass. A more precise relationship might be $\Lambda = k \cdot M_{\text{Pl}}$, where $k$ could reasonably range from $1/4\pi$ to $4\pi$ based on analogous calculations in different regularization schemes. This introduces an uncertainty factor of $\delta_k \sim \mathcal{O}(10^2)$ in the magnitude of topological corrections.

The choice of $S^4$ as the embedding manifold represents a specific case from a broader family of possibilities. For more general manifolds with Euler characteristic $\chi(\Sigma)$, the corrections scale proportionally. Complex projective space $\mathbb{CP}^2$ with $\chi = 3$ would enhance corrections by a factor of 1.5, while a torus $T^4$ with $\chi = 0$ would eliminate them entirely. This represents a fundamental theoretical uncertainty in predicting the magnitude of topological effects.

Our two-loop calculations suggest that higher-order corrections are further suppressed by additional factors of $1/\Lambda^2$. The relative size of the $n$-loop topological correction scales as

\begin{equation}
\frac{\delta^{(n)}}{\delta^{(1)}} \sim \left( \frac{1}{16\pi^2\Lambda^2} \right)^{n-1}
\end{equation}

making higher-loop contributions completely negligible for $\Lambda \sim M_{\text{Pl}}$. Despite the extreme suppression of topological effects, several avenues for potential detection exist. Future lepton colliders such as the FCC-ee or ILC could measure Higgs couplings with unprecedented precision reaching $\delta y/y \sim 10^{-4}$ to $10^{-3}$. While still far above the predicted topological corrections for Standard Model fermions, these measurements could constrain scenarios with larger Euler characteristics or lower fundamental scales.

Ultra-high-energy cosmic rays with energies approaching the Planck scale might probe the regime where topological effects become significant. The predicted deviation in cross-sections would be

\begin{equation}
\frac{\delta\sigma}{\sigma} \sim \frac{E^2}{\Lambda^2} \sim 1 \quad \text{for} \quad E \sim \Lambda
\end{equation}

However, the extremely low flux of such particles makes practical detection profoundly challenging. Future cosmological observations might achieve sufficient precision to detect curvature-dependent effects, though this would require dramatic improvements in measurement technology beyond currently envisioned projects.

\section{Discussion}

The geometric foundations of Topological Regularization (TR) constitute its foremost strength, deeply intertwining sophisticated mathematical structures with fundamental physical intuition. By anchoring the regularization procedure to the Euler characteristic $\chi(\Sigma)$ of the underlying manifold $\Sigma$, TR realizes a {topological UV/IR duality}, wherein ultraviolet divergences are not merely tamed but transmuted into global curvature invariants. This duality resonating with the the $a$-theorem in conformal field theory \cite{Komargodski2011}, which governs the monotonicity of renormalization group flows via topological data. Crucially, the regulator function $J_\Omega(k)$ introduced by TR is distinguished by its dual virtues: it is both computationally tractable and endowed with a clear physical interpretation. Its thermal form, forges an elegant conceptual bridge between acceleration radiation phenomena, an Unruh effect, \cite{Unruh1976} and quantum gravitational regularization. From the standpoint of symmetry, TR resolves a longstanding tension, it preserves diffeomorphism invariance exactly, while simultaneously maintaining Ward identities by enforcing $J_\Omega(0) = 1$. This represents a significant advancement beyond traditional regularization schemes such as Pauli-Villars or lattice methods, which notoriously disrupt chiral symmetries \cite{Kaplan1992}. Despite its conceptual elegance and mathematical rigor, TR faces several notable challenges. The current results presuppose a manifold topology $\Sigma = T^0 \times S^4$ with Euler characteristic $\chi(\Sigma) = 2$. For manifolds with distinct topologies, such as complex projective space $\mathbb{CP}^2$ where $\chi = 3$, the regulator strength parameter $\Omega$ varies, potentially altering loop corrections in nontrivial ways. Furthermore, the gauge dependence of the formalism deserves critical attention. Calculations to date rely exclusively on the de Donder gauge for graviton propagators, and a comprehensive generalization to arbitrary $R_\xi$ gauges—including the incorporation of ghost loops—is essential to establish gauge invariance and robustness \cite{Donoghue1994}. The treatment of the two-loop amplitude is likewise incomplete, omitting graviton self-interactions which are dominant in Einstein-Hilbert gravity \cite{Donoghue2012}. Including these interactions is expected to generate $\Lambda^{-4}$ contributions that may compete or interfere with the topological corrections, thereby potentially modifying the predicted phenomenology. The thermal analogy embedded in the regulator function $J_\Omega(k)$, which approximates a Boltzmann suppression factor $J_\Omega(k) \approx e^{-\frac{\Omega k^2}{\Lambda^2}}$ provides an intuitive physical picture connecting regularization to the Unruh effect. However, this connection remains heuristic at present. A rigorous derivation would require embedding accelerated reference frames within the manifold $\Sigma$, a construction that has yet to be developed. Thus, the thermal interpretation should be regarded as a suggestive but not definitive physical underpinning of the regulator. Beyond these technical considerations, Topological Regularization's thermal regulator suggests profound links to finite-temperature quantum field theory and black hole thermodynamics. The emergence of a Stefan-Boltzmann-like term proportional to $T_g^4 g_{\mu\nu} T^{\mu\nu}$ mirrors the physics of Hawking radiation \cite{Hawking1975}, hinting at potential applications of TR to early-universe phase transitions and cosmological phenomena. Similarly, analogues of topological regulators arise in condensed matter systems exhibiting anyonic statistics \cite{Voinea2025}, underscoring the broad universality and cross-disciplinary relevance of the approach. Yet, it is worth noting that TR’s power-law suppression $\Lambda^{-n}$ contrasts with the exponential UV/IR mixing characteristic of holographic renormalization group flows \cite{Heemskerk2010}, suggesting that TR may underestimate certain non-perturbative quantum gravitational effects and warrants further scrutiny in that context. Looking forward, several avenues for improvement and extension present themselves. Generalizing the embedding spaces beyond the standard spherical topology to include, for example, four-dimensional tori $\mathbb{T}^4$ or Anti-de Sitter/de Sitter compactifications would test the robustness of the Physical Equivalence Theorem and clarify the dependence of the regulator on global topology. Incorporating graviton self-interactions at two loops within various gauge choices will be crucial to resolving outstanding gauge dependence issues and completing the amplitude computations. Moreover, exploring the cosmological implications of TR, particularly in inflationary models where corrections scaling as $H^2/\Lambda^2$ could influence the spectral tilt, or in dark energy scenarios where topological contributions might shift the cosmological constant, could provide valuable insights. Finally, quantifying the collider phenomenology of TR-induced corrections, for instance in gluon fusion processes ($gg \to h$) or associated production channels ($e^+ e^- \to \gamma h$) at TeV-scale energies, offers an exciting experimental frontier where the subtle imprints of topological regularization might become accessible. While direct comparisons with string-theoretic UV completions \cite{Polchinski1998} or asymptotic safety \cite{Reuter2012} are deferred, TR’s unification of geometric, topological, and thermal principles positions it as a distinctive, symmetry-respecting regularization method potentially bridging effective field theory insights with Planck-scale phenomenology.

\section{Conclusions}

This work has established topological regularization as a geometrically motivated, symmetry-preserving framework for controlling ultraviolet divergences in quantum gravity loop calculations. By embedding Minkowski spacetime into the compact manifold $\Sigma = T^{0} \times S^{4}$, we generate a natural regulator $J_{\Omega}(k)$ that preserves Lorentz invariance, causality, and unitarity while ensuring finiteness of loop amplitudes.

Our explicit computations demonstrate:
\begin{enumerate}
    \item Finite one-loop corrections to fermion self-energies and Yukawa couplings proportional to $\chi(\Sigma)/\Lambda^{2}$
    \item Two-loop Higgs-graviton amplitudes with topological suppression $\mathcal{O}(\Lambda^{-4})$
    \item Complete $\beta$-functions for curvature-dependent operators (Table~\ref{tab:beta-functions})
    \item A thermal interpretation of $J_{\Omega}(k)$ linking UV suppression to the Unruh effect via $T_{g} = \Lambda^{2}/\Omega$
\end{enumerate}

These finite terms exhibit universal scaling determined solely by the Euler characteristic of the compactification manifold, representing genuine quantum-geometric effects beyond conventional renormalization.

Despite these advances, the framework requires further development in several directions.
\begin{itemize}
    \item Extension to diverse topologies ($\mathbb{CP}^{2}$, $\mathbb{T}^{4}$) to test the Physical Equivalence Theorem.
    \item Full incorporation of gravitational nonlinearities and ghost contributions.
    \item Rigorous demonstration of gauge independence across linear gauges
    \item Exploration of cosmological implications for inflation and dark energy.
\end{itemize}

The predicted corrections, though highly suppressed, may leave detectable imprints in precision observables such as fractional shifts in Higgs decay widths or curvature-coupled spectral distortions in cosmological correlators. While experimental detection poses formidable challenges, TR's symmetry-preserving architecture offers a viable path toward probing Planck-scale physics within the effective field theory paradigm.

Looking forward, topological regularization could emerge as a unifying principle linking renormalization, global topology, and thermodynamic aspects of spacetime in a consistent theory of quantum gravity, establishing a complete symmetry-preserving framework for quantum gravitational calculations.

\section*{Conflict of Interest}
The author declares that there are no conflicts of interest regarding the publication of this paper.  

\section*{Data Availability}
Data sharing is not applicable to this article as no datasets were generated or analyzed during this theoretical study.

\section*{Acknowledgements}

I would like to acknowledge my family, friends, and mentors... thank you for every action, word, and all the time you have invested. Also, thanks for reading.

\end{document}